%% file: main.tex
\documentclass[a4paper,twoside]{article}
\newcounter{the_style}
\input{main.macros}
\title{On Variational Data Assimilation in Continuous Time}
\author{Jochen Br\"{o}cker\\
Max--Planck--Institut f\"{u}r Physik komplexer Systeme\\
N\"{o}thnitzer Strasse~34\\
01187 Dresden \\
Germany}
\begin{document}
\maketitle
\begin{abstract}
Variational data assimilation in continuous time is revisited. 
The central techniques applied in this paper are in part adopted from the theory of optimal nonlinear control.
Alternatively, the investigated approach can be considered as a continuous time generalisation of what is known as weakly constrained four dimensional variational assimilation (WC--4DVAR) in the geosciences. 
The technique allows to assimilate trajectories in the case of partial observations and in the presence of model error. 
Several mathematical aspects of the approach are studied. 
Computationally, it amounts to solving a two~point boundary value problem.
For imperfect models, the trade off between small dynamical error (i.e.~the trajectory obeys the model dynamics) and small observational error (i.e.~the trajectory closely follows the observations) is investigated.
For (nearly) perfect models, this trade off turns out to be (nearly) trivial in some sense, yet allowing for some dynamical error is shown to have positive effects even in this situation.
The presented formalism is dynamical in character; no assumptions need to be made about the presence (or absence) of dynamical or observational noise, let alone about their statistics.
\end{abstract}
\section{Introduction}
\label{sec:introduction}
Suppose we are given a time series
\beq{equ:10.10}
\{\eta_t \in \R^d, t \in [t_s, t_f]\}
\eeq
which we will refer to as {\em observations} or {\em measurements}\footnote{in this paper, italics are used to introduce technical terms}.
The time $t$ is assumed to be continuous. 
Note that the observations might be higher dimensional (i.e.\ $d \geq 1$ is permitted).
A {\em model} for the observations is usually defined by certain dynamical equations
\begin{eqnarray}
\dot{x}_t & = & f(x_t), \label{equ:10.20-1}\\
y_t & = & h(x_t), \label{equ:10.20-2}
\end{eqnarray}
where $x_t \in \R^D$ will be referred to as the {\em state}, and $y_t$ as the {\em output} of the model at time $t$.
Entire trajectories will be denoted by $\{x_t , t \in [t_s, t_f]\}$ or simply $\{x \}$ if the time interval is clear from the context.
The problem considered in this paper is to make the devations between model output and observations small in the interval $[t_s, t_f]$, or in other words, to find trajectories $\{x_t, t \in [t_s, t_f] \}$ so that $y_t \cong \eta_t$ for all $t$. 
Different communities use different names for this type of problem.
In the geophysical sciences, it is mostly referred to as ``data assimilation''.
Making this notion more precise obviously requires to quantify the term ``deviation''.
Mainly for computational reasons, it is popular to work with a quadratic measure of deviation, that is,
\beq{equ:10.30}
A_T := \frac{1}{2}\int_{t_s}^{t_f} |y_t - \eta_t |^2 \dd t,
\eeq
or similar.
Obviously, $A_T$, henceforth called the {\em tracking error}, depends not only on the model as specified in Equations~(\ref{equ:10.20-1}-\ref{equ:10.20-2}), but also on the initial condition $x_{s}$.
A few words as to the relevance of this problem might be in order.
Firstly, we might be interested in the trajectory $\{x\}$ for diagnostical purposes. 
If $\{h(x)\}$ reproduces the observations well, we might reasonably hope that $\{x\}$ holds clues about the ``underlying state'' of the observed system, such as regime changes.
At least, we might learn something about the quality of our model equations.
Secondly, the eventual goal might be to forecast future values of either observed or even unobserved states. 
If proper initial conditions are at hand, forecasts are generated by running the dynamical model~\eqref{equ:10.20-1} into the future. 
Data assimilation provides a means to find proper initial conditions. 
At first sight, it seems reasonable to try the following variational approach to the problem: Choose the trajectory $\{x\}$ of the model~(\ref{equ:10.20-1},\ref{equ:10.20-2}) so that the output $\{y\}$ minimizes the tracking error $A_T$. 
A more in~depth analysis though would reveal that, by following this route, we are very likely to run into numerical difficulties, since the problem we are facing has a very poor condition.
This is looked at in detail in Section~\ref{subsec:dynamical-perturbations-motivation}, but roughly speaking, a poor condition means that small changes in the problem specification (e.g.\ small changes in $\{ \eta \}$ or the model $f$) can entail very large changes in the solution $\{ x \}$.
Almost all numerical algorithms though work by solving a series of nearby problems.
If these small changes cause large changes in the corresponding solutions, the algorithm will have trouble to converge.
As an aside, this problem is not restricted to instable systems, as we will see. 
A stable system creates essentially the same problems as an instable system. 
If instability was really the only cause of trouble, a reversal of time would solve the issue. 
Notwithstanding these concerns, the abovementioned approach has been considered in a large number of publications.
In the geosciences, this data assimilation technique is known as (strongly constrained) four dimensional variational data assimilarion, or 4D--VAR;
Several dynamical weather forecasting systems employ this approach operationally, and certainly with success.
For a discussion of the approach and applications to geophysical systems see for example~\citet{ledimet86,talagrand87,courtier87,rabier93,pires96,apte08}, a list which is by no means exhaustive.
In~\citet{FAS,broecker01,ridout02}, a very similar approach was considered (at least as a motivation), albeit from a nonlinear dynamics' perspective; furthermore, different solution strategies were employed. 
Several interesting observations were reported.
These observations allow for an interpretation within the framework of the present paper, to be discussed in Section~\ref{sec:other-work}.
This discussion might also pertain (at least partly) to operational implementations of 4D--VAR, and hopefully explain why the strongly constrained 4D--Var approach works to some extent, despite the abovementioned reservations.
The technique considered in this paper is motivated by similar investigations in the engineering literature~\citep[see for example][]{sage68,JAZ,sontag98}, as well as a recent paper by \citet{judd08-2}. 
In the geosciences, the shortcomings of 4D--VAR lead to the developement of weakly constrained~(WC) 4D--VAR, which might be understood as a special case of the following approach \citep[see for example][]{derber89,tremolet06,apte08}.
Consider a modified model 
\begin{eqnarray}
\dot{x}_t & = & f(x_t, u_t), \label{equ:10.110-1} \\
y_t & = & h(x_t) \label{equ:10.110-2},
\end{eqnarray}
which allows for a dynamical perturbation $u_t$, with the idea that $f(x_t, 0)$, the {\em free running system}, constitutes our ideal model.
This means that we expect truth to behave essentially as described by the vector field $f(\cdot, 0)$, which encompasses all our \textsc{a priori} understanding of the problem at hand.
In particular, $f(\cdot, 0)$ is the model we might want to use in order to extrapolate trajectories into the future.
In the theory of optimal control~\citep{sage68}, $u_t$ plays the role of a control parameter which allows to be manipulated externally, in order to force either the state $x_t$ or the output $\eta_t$ to exhibit some desired behaviour.
The word ``control'' though has an entirely different meaning in geophysical modeling.
Further to that, the interpretation of $u_t$ in the present problem is different, and more that of a model imperfection or error, whence the word ``dynamical perturbation'' was chosen here for $u_t$.
This interpretation might suggest that the existence of a perfect model is assumed, and that in the (near) perfect model case, $u_t$ is not needed.
The presented methodology however has no need to assume the existence of a perfect or even near perfect model.
Furthermore, it will become clear that there are good reasons to allow for some dynamical perturbation $u_t$ even if the free running model $f(\cdot, 0)$ has in fact generated the data (perfect model case). 
Whatever the interpretation of $u_t$, we are striving to reproduce the observations without too large deviations from free running model $f(\cdot, 0)$, that is, without too large dynamical perturbations.
Hence the variational approach is modified by adding a further term to $A_T$, referred to as {\em modelling error}, penalizing large dynamical perturbations, that is, large deviations from the ideal model $f(\cdot, 0)$.
More specifically, we consider the {\em action integral} 
\begin{eqnarray}
A_{\alpha} & = & (1 - \alpha) A_T + \alpha A_M \nonumber \\
 & = & \frac{(1 - \alpha)}{2} \int_{t_s}^{t_f}(\eta_t - y_t)^T R(\eta_t - y_t) \dd t + \frac{\alpha}{2} \int_{t_s}^{t_f} u_t^T S u_t \dd t.
\label{equ:10.120}
\end{eqnarray}
Here $\alpha \in [0, 1]$ is a weighting parameter.
Varying $\alpha$ allows to explore the trade--off between the desired goals of minimizing the tracking error and keeping the dynamical perturbations small.
The positive definite matrices $R$ and $S$ help to normalize the different components of observational and dynamical errors in $A_T$ and $A_M$ to about the same orders of magnitude. 
The factor $1/2$ simplifies subsequent expressions.
Several authors have interpreted $R$ and $S$, respectively, as (the inverse of) observational and dynamical error covariances.
More generally, an action integral of the form~\eqref{equ:10.120} can be motivated as a maximum--likelihood--approach to the smoothing problem~\citep[see][, Chap.5, Sec.3 for details]{JAZ}.
This interpretation though relies on a large number of assumptions which are hard to verify or even outright wrong.
At the same time, it is not clear what the benefits of such a stochastic interpretation would be in the present context.
In fact, other reasonable options for~$A_T$ and~$A_M$ are conceivable which would not have an immediate interpretation in stochastic terms.
For example, setting 
\beq{equ:10.35}
A_M = \int_{t_s}^{t_f} \left( f\left(x_t, u_t \right) - f\left(x_t, 0 \right) \right)^T S \left(f\left(x_t, u_t \right) - f\left(x_t, 0 \right) \right) \dd t
\eeq
would directly measure the deviations from the ideal model.
In addition, $R$ and $S$ could depend on $t$, too, thereby weighting different observations and dynamical perturbations differently (e.g.\ if observations from the remote past are deemed less important than more recent ones). 
Further examples are considered in Section~\ref{sec:other-work}.
This introductory section will finish with an overview over the following sections.
The next section recalls the basic facts from the calculus of variations, deriving a necessary condition for~(\ref{equ:10.120}) to have a minimum under the constraints~(\ref{equ:10.110-1},\ref{equ:10.110-2}).
This necessary condition has the form of a Hamiltonian two~point boundary value problem.
In Section~\ref{sec:lagrange-multiplier}, properties of the solution are discussed. 
In particular, an interpretation of the Lagrange multiplier is given. 
It turns out that the Lagrange multiplier conveys important information as to the stability of the solution.
Section~\ref{sec:solving-bvps} briefly explains approaches to solve the two~point boundary value problem as well as other technical aspects of the approach.
Most importantly, the concept of continuation is discussed.
In Setion~\ref{sec:other-work}, some related work is discussed, in particular the connection between the present paper and a recent work by~\citet{judd08-2}.
Finally, Setion~\ref{sec:numerical-observations} discusses observations from a numerical experiment.
\section{Necessary conditions for an optimum}
\label{sec:necessary-conditions}
The reader is reminded of some of the basic facts from variational calculus~\citep[for details, see e.g.][]{sage68}. 
Let the {\em Lagrangian} $L(x, u, t)$ be a function of time $t$ and the variables $x$ and $u$. 
The task is to minimize the action integral
\beq{equ:10.40}
A(\{x, u\}) := \int_{t_s}^{t_f} L(x_t, u_t, t) \dd t
\eeq
over all trajectories $\{x, u\}$ that satisfy a dynamical constraint of the form 
\beq{equ:10.50}
\dot{x}_t = f(x_t, u_t)
\eeq
and maybe the initial condition
\beq{equ:10.60}
x_{t_s} = \xi,
\eeq
where $\xi$ is given.
In this paper, Lagrangians will mostly be of the form
\beq{equ:10.65}
L = (\eta_t - h(x))^T R (\eta_t - h(x)) + \frac{1}{2} u^T S u
\eeq
or similar.
We will consider both the case of a specified initial condition and the case where the initial condition is free and hence subject to optimisation.
These two cases will be referred to as fixed initial condition problem and free initial condition problem, respectively. 
Any trajectory $\{x, u\}$ that satisfies Equation~\eqref{equ:10.50} (and also~\eqref{equ:10.60} in the fixed initial condition problem) is called a {\em feasible} trajectory.
Let $\{\hat{x}_t, \hat{u}_t\}$ denote a minimizing trajectory, that is, a feasible trajectory with the property that $A(\{\hat{x}_t, \hat{u}_t\}) \leq A(\{x, u\})$ for any other feasible trajectory $\{x, u\}$.
By $\hat{A} = A(\{\hat{x}_t, \hat{u}_t\})$, we will denote the optimum value of the action.
In order to formulate a necessary condition for $\{\hat{x}_t, \hat{u}_t\}$, define the {\em Hamiltonian}
\beq{equ:10.70}
H(x, u, \lambda, t) := L(x, u, t) + \lambda^T f(x, u).
\eeq
Then for $\{\hat{x}_t, \hat{u}_t\}$ to be a minimizing trajectory, it is necessary that $\{\hat{x}_t, \hat{u}_t, \hat{\lambda}_t\}$ is a stationary point of the {\em augmented action}
\begin{eqnarray}
\bar{A}(\{x, u, \lambda\}) 
& := & \int_{t_s}^{t_f} H(x_t, u_t, \lambda_t, t) 
	- \lambda^T_t \dot{x}_t \nonumber \\
& = & \int_{t_s}^{t_f} L(x_t, u_t, t) 
	- \lambda^T_t \left[ \dot{x}_t - f(x_t, u_t) \right] \dd t. \label{equ:10.80}
\end{eqnarray}
Using the calculus of variations, it can be shown that in order for $\{\hat{x}_t, \hat{u}_t, \hat{\lambda}_t\}$ to be a stationary point of the augmented action, it is necessary that they solve the two~point boundary value problem
\begin{eqnarray}
\pdq{H_t}{x} + \dot{\lambda}_t & = & 0, \label{equ:10.90-1} \\
\pdq{H_t}{\lambda} - \dot{x}_t & = & 0, \label{equ:10.90-2} \\
\pdq{H_t}{u} & = & 0, \label{equ:10.90-2.5} \\
\mathrm{(a)} \qquad \lambda_{t_s} = 0, 
	&& \mathrm{(b)} \qquad x_{t_s} = \xi, \label{equ:10.90-3} \\
\lambda_{t_f} & = & 0.\label{equ:10.90-4}
\end{eqnarray}
Here, initial condition~(\ref{equ:10.90-3}a) holds for the problem with free initial condition, while~(\ref{equ:10.90-3}b) holds for a fixed initial condition.
The corresponding boundary conditions for problems with either both ends or even only the final value of $\{ x \}$ specified should be evident.
Sometimes, instead of boundary conditions like~(\ref{equ:10.90-3},\ref{equ:10.90-4}), additional penalty terms of the form
\beq{equ:background}
A_S := (x_{t_s} - \xi_s)^T B_s (x_{t_s} - \xi_s),
\qquad 
A_F := (x_{t_f} - \xi_f)^T B_f (x_{t_f} - \xi_f)
\eeq
appear in the action integral. 
In control theory, such terms are referred to as start or end point penalties, while in data assimilation the term $A_S$ is called {\em background error}.
If present, such terms lead to boundary conditions of the form
\beq{equ:background_bc}
\lambda_{t_s} = B_s (x_{t_s} - \xi_s),
\qquad 
\lambda_{t_f} = B_f (x_{t_f} - \xi_f).
\eeq
Inserting the definition of the Hamiltonian into Equation~\eqref{equ:10.90-2} gives back the original dynamical condition~\eqref{equ:10.20-1}, while Equation~\eqref{equ:10.90-1} can be written as 
\[
-\dot{\lambda}_t = \left(\pdq{f_t}{x}\right)^T\lambda_t +  \left(\pdq{L_t}{x}\right)^T.
\] 
The coupling between the dynamical relations~\eqref{equ:10.90-1} and~\eqref{equ:10.90-2} is furnished by the static condition~\eqref{equ:10.90-2.5}.
The role of this condition becomes much more prominent if further inequality conditions are imposed on the dynamical perturbations, in which case the celebrated Pontryagin Maximum Principle~\citep[see for example][]{sontag98,sage68} states that condition~\eqref{equ:10.90-2.5} is to be replaced by 
\beq{equ:10.95}
	H(x_t, u_t, \lambda_t, t) \leq H(x_t, w, \lambda_t, t)
\eeq 
for all feasible perturbations $w$.
If there are no further inequality constrains on the dynamical perturbations, the Maximum Principle obviously agrees with condition~\eqref{equ:10.90-2.5}.
It is very convenient if the problem is set up in a way that allows for condition~\eqref{equ:10.90-2.5} to be solved for $u_t$, which can then be eliminated from the entire problem.
To simplify the subsequent discussion, we will from now on assume this situation.
The vector field
\beq{equ:10.100}
\mathcal{H}(\lambda, x) = %
\left( -\pdq{H}{x}, \pdq{H}{\lambda} \right)(\lambda, x)
\eeq 
is called the {\em Hamiltonian vector field}, where it is understood that on the right hand side, the dynamical perturbation $u$ has been expressed as a function of $\lambda$ and $x$, using the condition~\eqref{equ:10.90-2.5}. 
Although the stationarity conditions~(\ref{equ:10.90-1}-\ref{equ:10.90-4}) are but necessary conditions for a minimum of the action, they provide us with a practical means to calculate solution candidates.
What we have gained is that the minimisation under a $D$--dimensional differential equation as constraint has been replaced by a $2D$--dimensional differential equation.
Note however that this differential equation is {\em not} an initial value problem, but rather a two~point boundary value problem. 
In an initial value problem, the entire state $(x_t, \lambda_t)$ would be specified at $t = t_s$, while in the present two~point boundary value problem,  the state is specified in part at the beginning and in part at the end of the time window.
We will finish this section with deriving the Hamiltonian equations for the case we are mainly interested in, which is a quadratic Lagrangian as in Equation~\eqref{equ:10.120} and a dynamical system of the form
\beq{equ:10.130}
\dot{x}_t = f(x_t) + u_t, \qquad y_t = C x_t, 
\eeq
That is, perturbations to all degrees of freedom, and linear output. 
The Hamiltonian equations for this setup read as 
\begin{eqnarray}
\dot{\lambda}_t & = & 
	-\DD f(x_t)^T \lambda_t
	+ C^T R \left( \eta_t - C x_t \right), \label{equ:10.140-1} \\
\dot{x}_t & = & f(x_t) + u_t, \label{equ:10.140-2} \\
u_t & = & -\frac{1 - \alpha}{\alpha} S^{-1} \lambda_t \label{equ:10.140-3}
\end{eqnarray}
The following facts are easily derived from the system~(\ref{equ:10.140-1}-\ref{equ:10.140-3}).
Firstly, if there is a trajectory $\{ x \}$ of the free running system (i.e.\ with $u_t = 0$) so that $C x_t = \eta_t$ for all $t$, then this trajectory is also a solution of~(\ref{equ:10.140-1}-\ref{equ:10.140-3}) with free initial condition.
The corresponding $\lambda_t$ vanishes.
In other words, in the perfect model scenario, we get the desired solution.
Secondly, the boundary condition~(\ref{equ:10.90-4}) implies that the dynamical perturbation vanishes at $t_f$.
This has implications for the forecasting problem. 
Unless some way is found to extrapolate the dynamical perturbation $u_t$ into the future (i.e.\ beyond $t_f$), forecasts for $\eta_t$ and $x_t$ are generated using the free running model $f(x)$ in~\eqref{equ:10.130}.
The fact that $u_{t_f} = 0$ means a smooth transition from the reconstructed orbit $\{x_t, t \in [t_s, t_f]\}$ to the forecast orbit $\{x_t, t \geq t_f\}$.
This feature presumably yields more realistic trajectories than if the dynamical perturbations were switched off suddenly.
Thirdly, if $\{ \lambda, x \}$ is a solution of~(\ref{equ:10.140-1}-\ref{equ:10.140-3}), then $\{ c \lambda, x \}$ is also a solution but with $c R, c S$ replacing the matrices $R$ and $S$, where $c \in \R$.
In fact, this is true in general: If $c L$ replaces the original Lagrangian $L$, then $\{ \lambda, x \}$ has to be replaced with $\{ c \lambda, x \}$.
Finally, we see from Equation~(\ref{equ:10.140-3}) that the case of no dynamical
perturbations, in other words, the ``naive'' minimisation problem considered at
the very beginning of this paper, in which $A_T$ (Equ.~\ref{equ:10.30}) is
minimized subject to the dynamical system
Equation~(\ref{equ:10.20-1}-\ref{equ:10.20-2}), is equivalent to taking the
limit $\alpha \to 1$.
Note that this corresponds to assigning a strong weight to the modelling error $A_M$. 
In this situation, there is only unidirectional coupling in~\eqref{equ:10.140-1}, namely from $x_t$ to $\lambda_t$.
This has led several investigators to ignore $\lambda_t$.
As will be seen in the next section though, $\lambda_t$ conveys interesting information in any optimisation problem under constraints.
%
%
%
%
%
%
\section{Lagrange multipliers and dynamical perturbations}
\label{sec:lagrange-multiplier}
A central motivation for the presented approach was that it allows for a regularisation of the problem, even in a situation in which the proposed model has actually generated the data. 
This aspect is looked at in detail in Subsection~\ref{subsec:dynamical-perturbations-motivation}, after a few general facts on Lagrange multipliers have been presented in Subsection~\ref{subsec:interpretation-lambda}.
\subsection{Interpretation of the Lagrange multiplier}
\label{subsec:interpretation-lambda}
The Lagrange multipliers are not just a mere byproduct of the presented approach but provide useful information.
In any optimisation problem under constraints, a Lagrange multiplier corresponding to a specific constraint describes the derivative of the objective function with respect to that constraint at the optimum.
This fact is a consequence of a somewhat more general result on parameter dependencies of the objective function at the optimum, of which we give an informal derivation.
By $\hat{A} $ and $\hat{\bar{A}}$, respectively, we denote the functions $A$ and $\bar{A}$ (i.e.\ the action and the augmented action, resp.) at the optimum.
Suppose the Hamiltonian $H$ in Equation~\eqref{equ:10.70} depends on an additional parameter $\theta$.
Assume that small but otherwise arbitrary perturbations are applied to both $\theta$ as well as the functions $\{x, u, \lambda\}$.
The total first order change in $\bar{A}$ is given by
\beq{equ:20.10}
\delta_{tot} \bar{A} = \delta \bar{A} + \frac{\partial \bar{A}}{\partial \theta} \delta \theta
\eeq
where $\delta \bar{A}$ denotes first order changes due to perturbations of $\{x, u, \lambda\}$.
If $\bar{A}$ is at optimum, then $\delta \bar{A} = 0$, so that in this situation, 
\beq{equ:20.20}
\delta_{tot} \bar{A} = \frac{\partial \bar{A}}{\partial \theta} \delta \theta
\eeq
From the definition of $\bar{A}$, it follows that at the optimum
\beq{equ:20.30}
\hat{\bar{A}} = \hat{A},
\eeq
identically in $\theta$, so that for the total derivatives
\beq{equ:20.32}
\frac{\dd \hat{\bar{A}}}{\dd \theta} = \frac{\dd \hat{A}}{\dd \theta}.
\eeq
Combining this with Equation~\eqref{equ:20.20}, we obtain
\beq{equ:20.35}
\frac{\dd \hat{A}}{\dd \theta} = \frac{\partial \hat{\bar{A}}}{\partial \theta}
\eeq
We apply this to the situation where the dynamics is given by 
\beq{equ:20.37}
\dot{x_t} = f(x_t, u_t, t) + \theta r_t
\eeq
with $r_t$ some function of time. 
Equation~\eqref{equ:20.35} in conjunction with~\eqref{equ:10.80} yields
\beq{equ:20.40}
\frac{\dd \hat{A}}{\dd \theta} (\theta = 0) = \int \lambda^T_t r_t \dd t, 
\eeq
demonstrating that the Lagrange multiplier describes the first order changes of the objective function at the optimum with respect to perturbations of the constraints (i.e.\ the dynamics). 
\subsection{Motivation for using dynamical perturbations}
\label{subsec:dynamical-perturbations-motivation}
In this subsection, the effect of dynamical perturbations on the tracking problem is investigated.
In a nutshell, it will turn out that without dynamical perturbations, relatively small changes of the dynamical system as well as of the observations can have relatively large effects on the optimal trajectory.
Allowing for dynamical perturbations will be seen to alleviate these effects.
For simplicity's sake, we will study an example that is in fact not a dynamical system.
Consider the problem
\beq{equ:50.10}
\mbox{Minimize } A(x) := \frac{1}{2} (x - \eta)^2 
\qquad \mbox{subject to } Mx = \delta, 
\eeq
where $x, z \in \R^n$, $M$ is an arbitrary $m \times n$~matrix, and 
$\delta \in \R^m$.
A ``dynamical perturbation'' $u$ (which is in fact static in this case) can be introduced into this problem by modifying it thus:
\beq{equ:50.30}
\mbox{Minimize } A(x, u) := \frac{1}{2} (x - \eta)^2 + \frac{1}{2} u^2 
\qquad \mbox{subject to } Mx = bu + \delta, 
\eeq
where $b$ is a scalar factor and $u \in \R^m$.
The solution to this problem is most conveniently described by letting $M = W^T S V$ be the singular value decomposition of $M$, with $W$ an $m \times m$ orthonormal matrix, $V$ an $n \times n$ orthonormal matrix, and $S$ an $m \times n$ matrix of the form
\[
S = \mtr{ccc|c}{%
\sigma_1 & & 0 & \\
 & \ddots &  & 0 \\
0 & & \sigma_m  & }, \qquad \sigma_i \geq 0.
\]
Introducing $x':= Vx$, $u':= Wu$, $\delta':= W\delta$, $\eta':= V\eta$ and using Lagrange multipliers, the problem becomes one of finding a stationary point of
\[
A(x, u, \lambda) = \frac{1}{2} (x - \eta)^2 + \frac{1}{2} u^2 - \lambda^T [Sx - bu - \delta].
\]
(Here we have switched back to our old notation, ommitting the dashes.)
It is straight forward to verify that
\begin{eqnarray}
x_i & = & \alt{ \frac{1}{b^2 + \sigma_i^2} \left(b^2 \eta_i + \sigma_i \delta_i \right), \qquad \mbox{if }\sigma_i > 0\\
\eta_i, \qquad \mbox{else} \label{equ:50.40-1}}\\
\lambda_i & = & \frac{1}{b^2 + \sigma_i^2} \left(\delta_i - \sigma_i \eta_i\right), \qquad i = 1\ldots m\label{equ:50.40-2}\\
\hat{A} & = & \frac{1}{2} \sum_{i = 1}^{m}
\frac{1}{b^2 + \sigma_i^2} \left( \sigma_i \eta_i - \delta_i \right)^2
\label{equ:50.40-3}
\end{eqnarray}
Obviously, $\lambda_i = \partial \hat{A} / \partial \delta_i$, as was proved in the last subsection.
We are interested in the effect of small changes in the data $\delta, \eta$ on the solution of problem~\eqref{equ:50.30}.
A serious difficulty that appears in data assimilation (and other contexts) is that the singular values $\sigma_i$ range over many orders of magnitude.
This can render the smaller singular values indistinguishable from zero.
Keeping this in mind, the following conclusions can be drawn from Equations~(\ref{equ:50.40-1}-\ref{equ:50.40-3}):
\begin{enumerate}
\item If $b > 0$, then $x, \lambda$ and $A$ change continuously if $\sigma_i \to 0$.
Furthermore, the derivatives of  $x, \lambda$ and $A$ with respect to $\delta, \eta$ remain finite.
\item Letting $b \to 0$, we have to assume that $\delta_i = 0$ whenever $\sigma_i = 0$.
We obtain
\begin{eqnarray}
x_i & = & \alt{ \frac{\delta_i}{\sigma_i}, \qquad \mbox{if } \sigma_i > 0\\
\eta_i \qquad \mbox{if } \sigma_i = 0, \delta_i = 0\label{equ:50.60-1}}\\
\lambda_i & = & \frac{1}{\sigma_i} \left(\eta_i - \frac{\delta_i}{\sigma_i} \right) \label{equ:50.60-2}\\
\hat{A} & = & \frac{1}{2} \sum_{i = 1}^{m} \left( \eta_i - \frac{\delta_i}{\sigma_i} \right)^2 \label{equ:50.60-3}
\end{eqnarray}
It is seen that the solution is discontinuous for $\sigma_i \to 0$.
Moreover, the influence of $\delta, \eta$ on the solution becomes infinite. 
If $\delta_i \neq 0$ for $\sigma_i = 0$, the problem becomes infeasible.
\end{enumerate}
It is worth pointing out that a nonzero $b$ does not bring the solution closer to any presumed true solution.
It merely ensures that small changes in the problem formulation result in small changes in the solution.
Numerical approaches to data assimilation work by solving a series of nearby problems of the form considered in this subsection.
If this series of nearby problems does not result in a series of nearby solutions, the approach will fail to converge.
These considerations motivate the introduction of dynamical perturbations $u$ not only if model error is presumed present, but even if the model has actually generated the data.
%
%
%
\section{Solving BVP's and continuation}
\label{sec:solving-bvps}
Two point boundary value problems (BVP's) have been thoroughly investigated, and a multitude of algorithms and solution techniques exist. 
Subsection~\ref{subsec:collocation-equations} contains a few words on solution techniques and points to some references on this subject. 
The Subsection~\ref{subsec:continuation} explains continuation, an important and useful technique when working with BVP's.
\subsection{BVP solution techniques} 
\label{subsec:collocation-equations}
Although the Hamiltonian BVP~(\ref{equ:10.90-1}-\ref{equ:10.90-4}) involves ordinary differential equations (ODE's), it is not an initial value problem, which is what most ODE solvers are intended for.
To solve BVP's, different strategies are needed.
For a comprehensive collection of various numerical techniques for solving BVP's, see~\citet{childs78}.
A general approach consists of approximating the flow over small time intervals to translate the original problem into a BVP in discrete time---essentially a large system of nonlinear equations.
For the moment, we are not using the fact that the vector field is Hamiltonian, so we will be assuming (until further notice) that the BVP is given as
\beq{equ:265}
\dot{z}_t = F(t, z_t), \qquad t \in [t_s, t_f], \qquad b(z_{t_s}, z_{t_f}) = 0,
\eeq
with $z_t \in \R^D$, $F$ a vector field on $\R^D$ and $b(., ..)$ a function representing the boundary conditions.
BVP solvers approximate $z_t$ at a series $t_s = t_0 < \ldots < t_N = t_f$ of temporal mesh points; we will write $\z := (z_{t_0}, \ldots, z_{t_N})$.
Usually, $\z$ is obtained by solving a set of equations
\beq{equ:270}
\Phi_i(\z) = 0, \qquad i = 0 \ldots N, 
\eeq
called {\em collocation equations}.
The collocation equations are effectively discrete time approximations of the original differential equation as well as the boundary conditions in~\eqref{equ:265}.
Note that the mesh points need not be equal to the time instances at which the observations $\{ \eta \}$ were sampled.
Typically, the collocation equations are solved using a quasi--Newton type algorithm, which involves a numerical approximation to the Jacobian of the collocation equations.
As an example for a BVP solver, the code \texttt{bvp4c}~by \citet{kierzenka01} implemented in Matlab uses a fourth order implicit Runge--Kutta scheme to approximate the differential relation.
The collocation approach was used for the numerical experiments in this paper.
This approach, albeit fast and reliable, becomes prohibitively expensive for larger systems, since the entire system of collocation equations is solved simultaneously. 
Even if the sparsity of the collocation equations is exploited, this approach is unlikely to work for relevant geophysical problems.
Hamiltonian BVP's though can be solved sequentially by the method of {\em invariant imbedding}, see for example~\citet{sage68}.
Whether this route yields a feasible approach for large scale systems will be subject to future research.
One might contemplate a completely different approach in which the original dynamics~(\ref{equ:10.110-1},\ref{equ:10.110-2}) are linearized, while the action integral is expanded to second order (if not already in this form).
The result is a so--called linear--quadratic programme.
The linear Hamiltonian equations obtained in this way though are {\em not} the same as the linearized original Hamiltonian equations.
In fact, this approach neglects important curvature terms and might even render a perfectly solvable problem infeasible.
This well known phenomenon is discussed in many textbooks on nonlinear programming. 
This approach is not recommended. 
\subsection{Continuation}
\label{subsec:continuation}
An important technique when dealing with BVP's (or indeed with any type of equation solving) is continuation.
Suppose the entire problem depends on a parameter $\alpha$. 
To every value of $\alpha$, there corresponds a solution $\{x(\alpha), 
\lambda(\alpha)\}$.
Sometimes a solution is easily obtained for a specific $\alpha_0$, while finding a solution at the desired value $\alpha_1$ is hard.
This situation calls for continuation, which means to first calculate a solution at the ``easy'' value $\alpha_0$ and then use that solution as an initialisation for a new run with a slightly altered $\alpha$, hoping that the solution for the new value of $\alpha$ is still reasonably close to the solution for $\alpha_0$.
This process is repeated, gradually moving $\alpha$ towards the desired value $\alpha_1$.
For continuation, it is most useful if the employed BVP solver accepts initial guesses in exactly the same format as it produces solutions, as is the case for \texttt{bvp4c}. 
In this paper, we will use continuation with respect to the mutual weighting between observational errors and dynamical errors.
To be specific, the action integral is written as $A_{\alpha} = (1 - \alpha) A_T + \alpha A_M$ with some fixed positive definite matrices $R, S$.
Starting with a very small model error penalty (i.e.\ $\alpha \cong 0$), the solution is obviously any trajectory $\{x, 
\lambda\}$  so that $h(x_t) \cong \eta_t$ (and $\lambda_t $ some large values) for all $t$.
Hence, $\alpha \cong 0$ appears to be a good starting value for continuation.
Another interesting option is continuation with respect to time. 
The interval $[t_s, t_f]$ can be divided into subintervals on which the BVP gets solved individually. 
Depending on which boundary conditions are used, either the dynamical perturbations or the orbit itself can be rendered smooth at the endpoints of the individual intervals. 
Splicing together the individual solutions should provide a fairly good initial guess for the optimal orbit on the entire interval.
Alternatively, a realtively short interval $[t_s, t_f]$ might be computed first.
With new measurements $\eta$ being aquired, the interval gets sequentially incremented.
Again, previous solutions projected into the future should provide a good initial guess for the updated solution.
%
%
%
%
%
%
\section{Relation to previous work}
\label{sec:other-work}
As was mentioned in the introduction, minimising the tracking error $A_T$ subject to the model~(\ref{equ:10.20-1},\ref{equ:10.20-2}) without dynamical perturbation was considered in several publications (outside the meteorological community), for example as a means for noise reduction by~\citet{FAS}.
Noise reduction refers to the following problem.
It is assumed that $\eta_t = \bar{x}_t + \xi_t$, where the dynamics of governing $\bar{x}_t$ are an identical copy of the model, and  $\xi_t$ is white noise.
In other words, the model is indeed perfect, up to observational errors.
Also, it is assumed that $C$ is the identity matrix, that is, there are no hidden state variables. 
In~\citet{FAS} and all other papers referred to in this subsection, the time is discrete.
In discrete time, application of standard Lagrange multiplier theory leads directly to the collocation equations. 
In order to keep the discussion as streamlined as possible, we will stay with continuous time here, tacitly applying the necessary modifications without any further note.
\citet{FAS} approach the Hamiltonian BVP using what the authors call ``manifold decomposition''.
This is an iterative procedure whereby the dynamics~\eqref{equ:10.20-1} is linearised about a trial solution and decomposed into the (linear) stable and unstable directions. 
Corrections to the trial solution are obtained by solving the linearized system forward in time along the stable and backward in time along the unstable directions. 
The Lagrange multipliers $\lambda_t$ are ignored. 
Several difficulties of this approach are readily acknowledged, most importantly, the bad condition of the problem, and that the solution is not in fact a solution of the as posed initially. 
The algorithm merely finds a solution of the model dynamics, starting at the observations and proceeding in steps which are guaranteed to be small.
In \citet{broecker01} and \citet{ridout02}, a very similar problem is discussed, the approach though is different from \citet{FAS}.
Both papers propose to minimize the {\em indeterminism}
\beq{equ:10.220}
I = \frac{1}{2} \int_{t_s}^{t_f} |\dot{x}_t - f(x_t)| \dd t,
\eeq
which, in the present terminology, is the modelling error $A_M$.
To minimize \eqref{equ:10.220}, gradient descent is used by~\citet{ridout02}, while~\citet{broecker01} employ a simplified Newton--Raphson strategy.
In all three works, the observations come in only as an initial trial solution for $\{x\}$ (recall that in both papers, $C x = x$). 
In subsequent iteration steps, the observations are ignored. 
This seems to be rather counterintuitive:
To minimise the tracking error $A_T$, you ignore the observations and minimize the dynamical error $A_M$ instead. 
Nonetheless, the approach appears to work well, at least if model error is absent.
Using the methodology presented in this paper, an interpretation of this somewhat curious finding of~\citet{ridout02,broecker01} is hazarded.
It seems that the solution strategy corresponds to a (somewhat crude) continuation scheme. 
Recall that a scheme without dynamical perturbation corresponds to a large model error penalty, or a large ratio $\frac{\alpha}{1 - \alpha}$.
In other words, the observations should eventually be ignored. 
Obviously, the observations cannot be ignored altogether. 
As was discussed, problems for large~$\alpha$ need to be solved using continuation, starting with small~$\alpha$.
The mentioned papers assume that full state information is available, hence the Equations~(\ref{equ:10.140-1} - \ref{equ:10.140-1}) can be written as
\begin{eqnarray}
\dot{\lambda}_t & = & 
	-\DD f(x_t)^T \lambda_t
	+ (1 - \alpha) \left( \eta_t - x_t \right), \label{equ:10.225-1} \\
\alpha \left( f(x_t) - \dot{x}_t \right) & = & \lambda_t, \label{equ:10.225-2}
\end{eqnarray}
The solution for small~$\alpha$ is $\{x, \lambda\} = \{\eta, 0\}$, exactly what is used by~\citet{broecker01,ridout02} to initialise their algorithms.
We can conclude that the solution strategy in these publications corresponds to a continuation scheme in which $\alpha$ is moved from $0$ to $1$, which is appropriate for small dynamical errors.
The remarks of this section might as well apply to (strongly constrained) 4D--VAR. 
The author certainly cannot claim to have sufficient knowledge of the specifics of 4D--VAR implementations, which certainly differ substantially from the algorithms used in the publications discussed in the present subsection.
Considering the discussed difficulties though that would be encountered if the strong constraint in 4D--VAR were taken ``too literally'', it is speculated that successful implementations use in fact some kind of weakened strong constraint, either with or without acknoledging this fact.
Therefore, some of the remarks of this section might as well pertain to (not so) strongly constrained 4D--VAR.
The ideas of~\citet{ridout02} were substantially expanded upon in a series of papers by Kevin~Judd~\citep{judd03,judd08,judd08-2}.
We will focus on~\citet{judd08-2}, as it developes ideas not unsimilar to the present setup. 
In that paper, is proposed to minimize the tracking error $A_T$ subject to
\beq{equ:10.230}
\dot{x}_t - f(x_t) = u_t
\qquad \mbox{and} \qquad A_M \leq \delta.
\eeq
Several options for the exact form of $A_T$ and $A_M$ are discussed. 
In particular for $A_M$, \citet{judd08-2} suggests either a mean quadratic error as considered here (Equ.~\ref{equ:10.30}) or a maximum deviation of the form $A_M = \max_t u_t^T S u_t$.
The first case is equivalent to minimizing an error functional $A = (1 - \alpha) A_T + \alpha A_M$ with an $\alpha$ depending on $\delta$. 
Hence this case is a special case of the situation considered in this paper.
Using a maximum deviation for $A_M$ is equivalent to imposing an inequality constraint of the form $u_t^T S u_t \leq \delta$,  which in continuous time requires applying the Pontryagin Maximum Principle, as was discussed in Section~\ref{sec:introduction}. 
As said, \citet{judd08-2} considers only discrete time, and the minimisation problem is solved using the standard Lagrange multiplier method.
The necessary conditions directly yield the collocation equations.
%
%
%
%
%
%
\section{Observations from a numerical experiment}
\label{sec:numerical-observations}
\subsection{A prop for reality, and the model}
\label{subsec:reality-model}
In this section, results from a few numerical experiments will be discussed.
As a prop for reality, consider 
\beq{equ:30.10}
\dot{X}_t = F(X_t) + \sigma s_t
\qquad
\eta_t = C X_t + \rho r_t,
\eeq 
where for the dynamics $F$ we will consider the Lorenz~'63 system as well as the Lorenz~'96 system. 
We will use a capital letters like $X_t$ for the reality, while $x_t$ is reserved for model trajectories.
The quantities~$s_t$ and~$r_t$ are random perturbations to the dynamics and the observations, respectively. 
Their exact form is discussed below. 
The model we plan to use for tracking $\eta_t$ will in both cases be of the form 
\beq{equ:30.30}
\dot{x}_t = f(x_t) + u_t, \qquad
y_t = C x_t,
\eeq 
where $f$ might be different from $F$.
The action integral $A_{\alpha}$ is as specified in Equation~\eqref{equ:10.120}, with $R$ and $S$ being diagonal matrices defined as follows.
Let $\var{x}$ be the temporal variance of some time series $\{x\}$.
If $\{x\}$ is vector valued, then $\var{x}$ is understood component--wise, with the individual components being denoted by $\var{x}_i$.
With this definition, we set 
\beq{equ:30.35}
R_{ii} = 1 / \var{Cx}_i, \qquad
S_{ii} = 1 / \var{f(x)}_i.
\eeq 
These constants are calculated off--line with $\{x\}$ being some typical trajectory of the model.
The random perturbations $r_t$ and~$s_t$ are supposed to account for model error.
They are often referred to as dynamical and observation error, respectively.
There is some sort of agreement in the community that taking these perturbations as white noise processes is an acceptable choice.
Notwithstanding the fact that in practice, model deficiencies are usually anything but white noise, we will essentially stay with this custom in order to obtain comparable results, albeit with some modifications necessitated by the present situation.
Using white noise in conjunction with dynamical systems in continuous time requires some caution.
It is well known that if a vector field is subjected to white noise perturbations, the standard theory of ordinary differential equations ceases to apply, and stochastic analysis has to be used. 
This concerns not only our prop for reality (Equ.~\ref{equ:30.10}), but the Hamiltonian equations~(\ref{equ:10.90-1}-\ref{equ:10.90-4}) as well, since they are subjected to the input $\{\eta\}$, which contains white noise components.
In practice though, the observed signal will always be discretely sampled.
The sampling mesh however need not be appropriate for the BVP--solver, which for numerical reasons might need a finer mesh.
In this situation, it becomes necessary to interpolate the observations at intermediate time points.
In order to guarantee that the interpolated observations are sufficiently regular, we will assume once and for all that $\{\eta\}$ has a spectrum limited to frequencies smaller in magnitude than some bandwidth $\beta$Hz.
This bandwidth has to be chosen as part of the data assimilation procedure and hence becomes a component of the model.
A way to ensure a bandwidth limited continuous signal $\{\eta\}$ is to sample the original observations with a rate of $2\beta$ and interpolate intermediate values.
Note that the approach is not only operationally feasible, but typically the only option in practice, as the observations are usually given as a set of discrete observations with fixed sampling rate. 
Solutions were computed using continuation with respect to $\alpha$, starting with $\alpha \cong 0$, gradually increasing $\alpha$ to the desired value. 
Here and also in the presentation of the results below, it proved useful to introduce the {\em inverse logit} $\gamma \in \R$ of $\alpha$ via $\gamma := \log(\alpha) - \log(1 - \alpha)$.
This transformation maps the unit interval monotonously onto the real line.
When speaking about ``plotting results versus $\alpha$'', we really mean plotting the results versus $\gamma(\alpha)$.
The quality of the solutions was evaluated in terms of the quadratic deviation to the real solution $X_t$, that is, 
\beq{equ:30.60}
A_A := \int_0^T (x_t - X_t) Q (x_t - X_t)^T \dd t,
\eeq
a quantity which we refer to as the {\em assimilation error}.
Here, $Q$ is given by a diagonal matrix with entries
\beq{equ:30.65}
Q_{ii} := 1 / \var{x}_i.
\eeq
It is worth noting though that in real world applications, $X_t$ is not known (or does not exist, depending on one's philosophical stance).
Hence, $A_A$ cannot be the basis for quality assessment under operational circumstances. 
In the situations studied here, $A_A$ was found to feature a unique minimum.
That such a minimum must exist is quite plausible. 
For $\alpha \to 0$, a vanishing penalty is imposed on the dynamical perturbations $u_t$, but a strong penalty on the observational error. 
The solution will follow the observations very closely, thereby modelling not only the observations themselves but also the observational noise. 
If dynamical noise is present, this requires a nonvanishing $u_t$, but this has no effect on the cost functional $A$.
Thus, the algorithm can use large~$u_t$ with impunity. 
A large~$u_t$ is however costly in terms of $A_A$, since the orbit $\{x\}$ will not be a solution of the free running model, and therefore, will deviate from $\{X\}$. 
On the other hand, for $\alpha \to 1$, the observations are ignored, and again, if dynamical noise is present, the solution will soon deviate from the true trajectory. 
These deviations have no effect on the cost functional~$A_{\alpha}$, but of course, they do have an adverse effect on the assimilation error $A_A$.
So somewhere in between, there should occur a minimum of $A_A$.
\subsection{Numerical results for Lorenz'63}
\label{subsec:numerical-results}
The Lorenz~'63 system is given by the vector field
\beq{equ:30.20}
f(x)
= \mtr{c}{%
	10 (x^{(2)} - x^{(1)}) \\
	28 x^{(1)} - x^{(2)} - x^{(1)} x^{(3)} \\
	x^{(1)} x^{(2)} - \frac{8}{3} x^{(3)}%
}.
\eeq 
In these experiments, we used $F = f$, but note that still a significant model error is present, due to the dynamical noise in the prop for reality.
The tracking problem was investigated for a selection of observational and dynamical noise amplitudes (see Tab.~\ref{tab:limit-vals}), and a few representative results are presented here.
The signal-to-noise ratio (SNR) is defined as the energy of the signal versus the energy of the noise. 
More specifically, we set $\sigma$ and $\rho$ as diagonal matrices so that
\beq{equ:30.40}
\SNR_D := 10 \log_{10} \frac{\var{F(X)}_i}{\sigma_{ii}^2}
\eeq
for the dynamical noise, and 
\beq{equ:30.50}
\SNR_O := 10 \log_{10} \frac{\var{\eta}}{\rho^2}
\eeq
for the observational noise. 
The noise amplitudes are defined with respect to some typical trajectory $\{X\}$ of the true dynamics.
The signal $\eta_t$ was sampled with 20Hz, the total assimilation window comprised 256~points, whence $t \in [0, 12.8]$.
Continuation was carried out in steps of $\gamma$ of not more than $.5$, commencing with $\gamma = -8$, which corresponds to $\alpha \cong \exp(-8)$.
In Figures~\ref{fig:am8-traj}--\ref{fig:a3p5-traj}, results are shown for one experiment with $\SNR_D = 5\dB, \SNR_O = 5\dB$.
In all figures, the three panels on the left hand side contain the three components of the original trajectory $\{ X \}$ over time (in grey) and the three components of the assimilated trajectory $\{x\}$ over time (in black).
The right hand side, upper panel, shows the three components of the dynamical perturbations over time. 
The $y$--axis has been scaled with $1 / \sqrt{S}$.
This means that dynamical perturbations of magnitude~$1$ in this plot are of similar amplitude as the model vector field.
The right hand side, lower panel, shows the observations (black narrow line with diamonds), the output $y_t = Cx_t$ (black wide line), as well as the noise--free observations $C X_t$ (grey wide line).
For illustrative purposes, this panel zooms out a subinterval of the entire time span.
Figures~\ref{fig:am8-traj}--\ref{fig:a3p5-traj} only differ in terms of the values of $\alpha$ used. 
Figure~\ref{fig:am8-traj} shows the results for $\gamma = -8$, which corresponds to $\alpha \cong 3.35 \cdot 10^{-4}$.
Figure~\ref{fig:a0-traj} shows the results for $\gamma = 0$, which corresponds to $\alpha \cong .5$.
This value for $\alpha$ gave the smallest assimilation error (see below).
Figure~\ref{fig:a3p5-traj} shows the results for $\gamma = 3.5$, which corresponds to $\alpha \cong 0.97$.
A few interesting facts emerge. 
As expected, the trajectories become smoother with increasing values of $\alpha$ as they follow the unperturbed model dynamics to a higher and higher degree.
The first component in particular starts off with following the noisy observations rather closely for small $\alpha$, becoming progressively smoother with increasing $\alpha$.
For $\alpha = .97$, the output looks essentially like that of an unperturbed Lorenz'63 system.
The plot of the dynamical perturbations (upper panel in the right hand column) shows decreasing amplitudes for increasing $\alpha$.
As was discussed earlier, we should expect the deviation between the true $\{X\}$ and the assimilated trajectory $\{x\}$ to be minimal for some $\alpha$.
This effect can be readily discerned from Figures~\ref{fig:am8-traj}--\ref{fig:a3p5-traj}.
For small $\alpha$ (Fig.~\ref{fig:am8-traj}), the assimilated trajectory is very irregular due to the observational noise being assimilated into the system.
For medium $\alpha$ (Fig.~\ref{fig:a0-traj}), the assimilated trajectory is quite regular and closely follows the true trajectory.
For large $\alpha$ (Fig.~\ref{fig:a3p5-traj}), the deviations start to increase again, apparently due to model error kicking in. 
This is evident for example at $t \cong 8.5$, where true $\{X\}$ and assimilated trajectory $\{x\}$ slip out of phase.
Figures~\ref{fig:A-dd3-do3}--\ref{fig:A-dd8-do6} display the various errors $A_A, A_M$, and~$A_T$.
Each figure corresponds to a specific pair of observational and dynamical noise strength (see figure captions and Table~\ref{tab:limit-vals}). 
In all figures, the assimilation error $A_A$ (Eq.~\ref{equ:30.60}) is plotted with a wide black line, the tracking error $A_T$ and the modeling error $A_M$ (Eq.~\ref{equ:10.120}) are plotted with a dashed line and a dash--dotted line, respectively.
The grey line represents the assimilation error in the first component only (denoted as $A_1$), namely the error between the model output and the observations without observational noise.
This line was also shifted by a constant to allow for comparison with the (total) assimilation error $A_A$.
As expected, the modeling error decreases with increasing $\alpha$, while the tracking error increases.
Across different experiments, these two quantities, albeit very similar, are not equal. 
For small $\alpha$, the modeling error becomes very large but settles off at a finite value.
The limiting values have been collected in Table~\ref{tab:limit-vals}.
\begin{table}[tb]
\begin{center}
\begin{tabular}{cccc}
Dyn.~Noise 	& Obs.~Noise 	& $A_M(\alpha = 0)$ & Figure \\[1ex]
3 					& 3						& 6			& \ref{fig:A-dd3-do3} \\
4 					& 4						& 5.5 	& \ref{fig:A-dd4-do4} \\
4 					& 6						& 3.2		& \ref{fig:A-dd4-do6} \\
5 					& 5						& 4.5 	& \ref{fig:A-dd5-do5} \\
6 					& 4						& 4.5		& \ref{fig:A-dd6-do4} \\
7 					& 5						& 3.5 	& \ref{fig:A-dd7-do5} \\
7 					& 7						& 2.5 	& \ref{fig:A-dd7-do7} \\
8 					& 6						& 3.4 	& \ref{fig:A-dd8-do6}
\end{tabular}
\caption{\label{tab:limit-vals}Investigated values for the dynamical and observational noise, limiting values for the modeling error $A_M$ for $\alpha \to 0$, and corresponding figures.}
\end{center}
\end{table}
The assimilation error shows a clear minimum in all cases.
The position of the minimum depends on both the dynamical and the observational noise strength. 
The presented and further numerical experiments indicate that the minimum of the assimilation error moves towards larger values of $\alpha$ with increasing observational noise, and towards smaller values with increasing dynamical noise.
The behaviour of the assimilation error of the first component $A_1$ (dotted line) is qualitatively similar to $A_A$.
In particular, the minimum of $A_1$ with respect to $\alpha$ is close to the minimum of $A_A$.
Of course, $A_A$ and $A_1$ would not be available under operational circumstances. 
The question arises whether the quantitative behaviour of $A_A$ can be predicted in advance, and in particular, whether the optimiser $\hat{\alpha}$ can be identified from properties of the model alone (plus some assumptions about the perturbations).
It has been suggested that for 4D--VAR, the weighting matrices $R$ and $S$, respectively, should be chosen as the inverse of the observational and dynamical error covariances, respectively.
It is hard to see how the error covariances could be determined in an independent fashion though (letting alone the question as to whether a nonlinear vector field with white noise perturbations is an appropriate model).
Therefore, appropriate weighting matrices $R$ and $S$ (or $\alpha$) need to be determined using some problem oriented criterion of solution quality.
Finding criteria is an important problem which requires further research. 
An operationally feasible approach is presented in~\cite{broecker09-4}.
The resulting weighting matrices $R$ and $S$ could be referred to as \textsc{a posteriori} effective error covariances.
\subsection{Numerical results for Lorenz'96}
\label{subsec:numerical-results-lor96}
Very similar experiments were carried out for the Lorenz'96 system. 
The Lorenz~'96 system has two time scales, the slow degrees of freedom $X_t \in R^K$ and the fast degrees of freedom $Z_t \in R^{K\cdot L}$.
The dynamics are given by 
\[
\dot{X}_t = F(X_t, Z_t), \qquad \dot{Z}_t = G(X_t, Z_t)
\]
with $F, G$ being defined as
\begin{eqnarray}
\label{equ:30.25-1}
F^{(k)}(x, z)
& = & x^{(k-1)} (x^{(k+1)} - x^{(k-2)}) - x^{(k)} + c - S^{(k)},\\
G^{l}(x, z) & = & -a_1 z^{(l+1)} (z^{(l+2)} - x^{(l-1)}) - a_2 z^{(l)} + x^{l/L},\\
S^{(k)} 
& = & \sum_{l = 0}^{L-1} z^{(k+l)},
\end{eqnarray}
where $k$ and $l$ are to be understood modulo $K$ and $K \cdot L$, respectively, with $K = 64$ and $L = 8$.  
Furthermore, $c = 18$ for all $k$, $a_1 = 100, a_2 = 10$.
The output was set to $\eta_t = CX_t + \rho r_t$, with $C_{i, j} = \delta_{2i, j}$, with $i = 1 \ldots K/2, j = 1 \ldots K$, that is, only every second slow variable was observed.
The sampling interval of the observations was~0.005.
Observational noise with standard deviation~$1$ was added, corresponding to a SNR~of around 14dB.
The entire time series comprised~2048 points, hence $t_f - t_s = 10.24$.
For assimilation, we use a model without the fast degrees of freedom, that is, 
\beq{equ:40.100}
f^{(k)}(x) = x^{(k-1)} (x^{(k+1)} - x^{(k-2)}) - x^{(k)} + c, 
\eeq
and the same $c$ as above.
As a BVP--solver, the NAG Fortran library routine D02RAF was used, which is largely based on the techniques described by~\citet{pereyra78}.
The results are shown in Figure~\ref{fig:lor96}.
This figure is to be interpreted as Figures~\ref{fig:A-dd3-do3}--\ref{fig:A-dd8-do6}  for the Lorenz'63 case.
Due to the wider variations of the quantities in the present case, the $y$--axis was scaled logarithmically.
Additionally, the modeling error in the hidden degrees of freedom was calculated, and is represented by the grey dash--dotted line.
For decreasing $\alpha$, this quantity does not increase as much as the overall modeling error $A_M$.
This is to be expected, since for small $\alpha$ the observed degrees of freedom are forced to follow the observations, which requires larger perturbations. 
In general, the behaviour of $A_M, A_T$, and $A_A$ is very similar to the Lorenz'63 experiments.
A notable exception is that the minimum of the overall assimilation error and the assimilation error of the observed components have their minimum in slightly  different places.
This suggests that in more complicated examples, more than one $\alpha$ is necessary. 
In other words, in the action integral $A_{\alpha}$ (Eq.~\ref{equ:10.120}), the matrices $R, S$ need to be regarded as tuning parameters and their influence on the solution investigated. 
This will be subject of a future study.
\input{figs}
\section{Conclusions}
\label{sec:conclusions}
In this paper, a variational approach to data assimilation in nonlinear dynamical systems was discussed. 
The necessary conditions for a minimum of the action integral can be written as a Hamiltonian two~point boundary value problem. 
It was argued that in order to improve the condition of this two~point boundary value problem, some form of regularisation is necessary.
One possible option is to allow for some deviations from the proposed dynamical model.
In analogy to related problems from control theory, these deviations could be termed ``controls'', but the word ``dynamical perturbations'' was used here.
The magnitude of the dynamical perturbations is included as an additional term in the action integral, and the problem becomes one of minimizing a combination of the deviation from the observation and the deviation from the proposed model.
The connection to several related papers was discussed, and the main findings in these papers were interpreted with in the framework of the present contribution.
In particular, it was discussed why several approaches in the literature work well, despite the fact that they do not allow for dynamical errors (which appears to contradict one of the main tenets of the present paper).
The solution is that, roughly speaking, the discussed algorithms in fact do allow for dynamical errors, although they do not appear explicitely in the problem formulation.
Furthermore, an explanation was given to the (probably paradoxical) finding of \citet{ridout02} and \citet{broecker01} that several algorithms work well despite the fact that they, after having been initialized properly, essentially ignore the observational data.
This has been explained here as a form of continuation.
Observations from a numerical experiment were reported. 
As a prop for reality, the Lorenz'63 and~'96 systems were considered, subject to both dynamical and observational error.
Observations of several (but not all) state variables of these systems were employed for data assimilation.
The effects of a finite sampling time and the necessity for smooting were discussed.
The problem of how to weight the tracking error versus the dynamical error in the action integral and the resulting effects on the solution was mentioned.
\bibliographystyle{plainnat}
\bibliography{/home/broecker/TeX/Literatur}
%
%
%
%
%
\end{document}

%% file: figs.tex
\begin{figure}
\begin{center}
\epsfig{file = 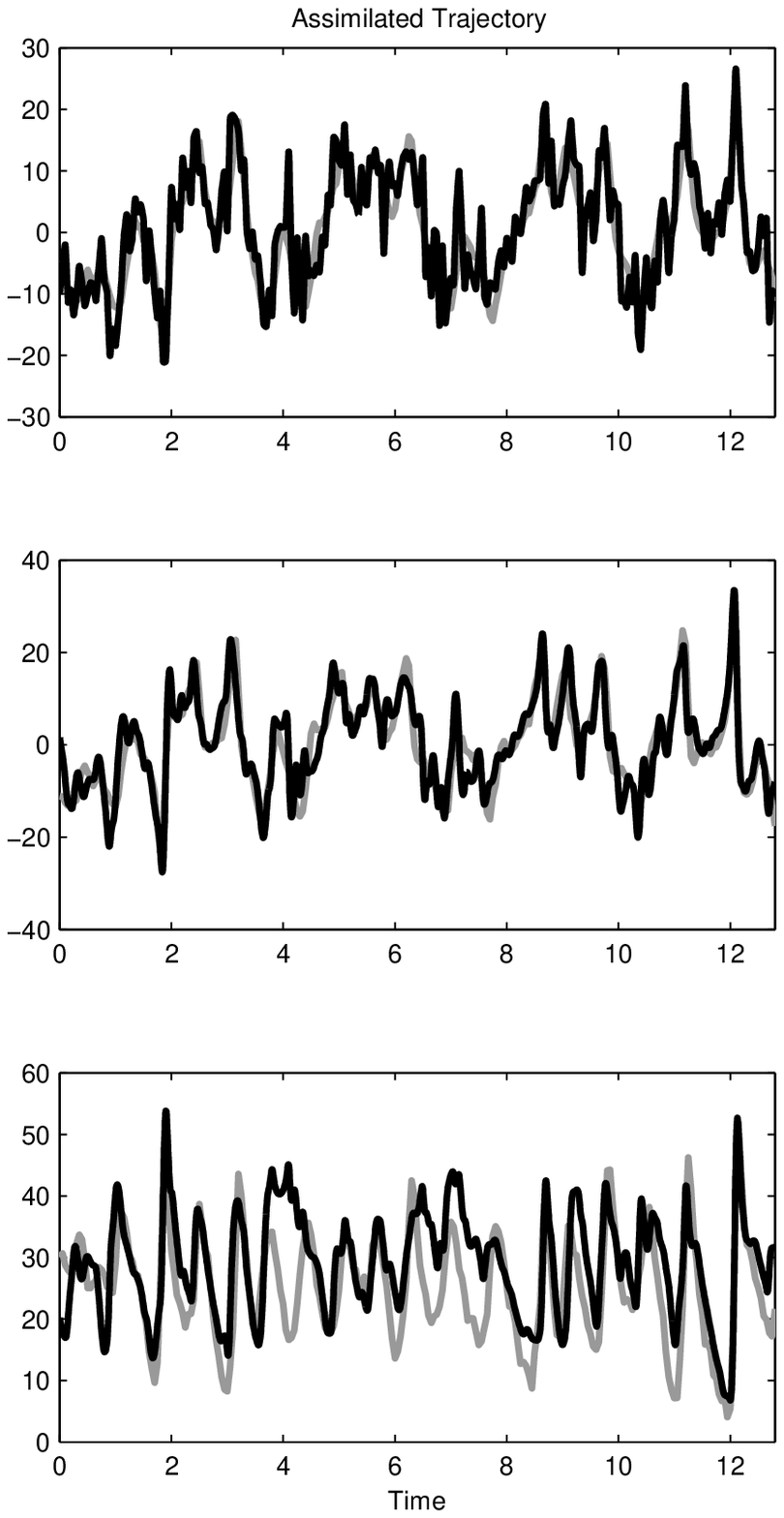, width = .49\textwidth}
\hfill
\epsfig{file = 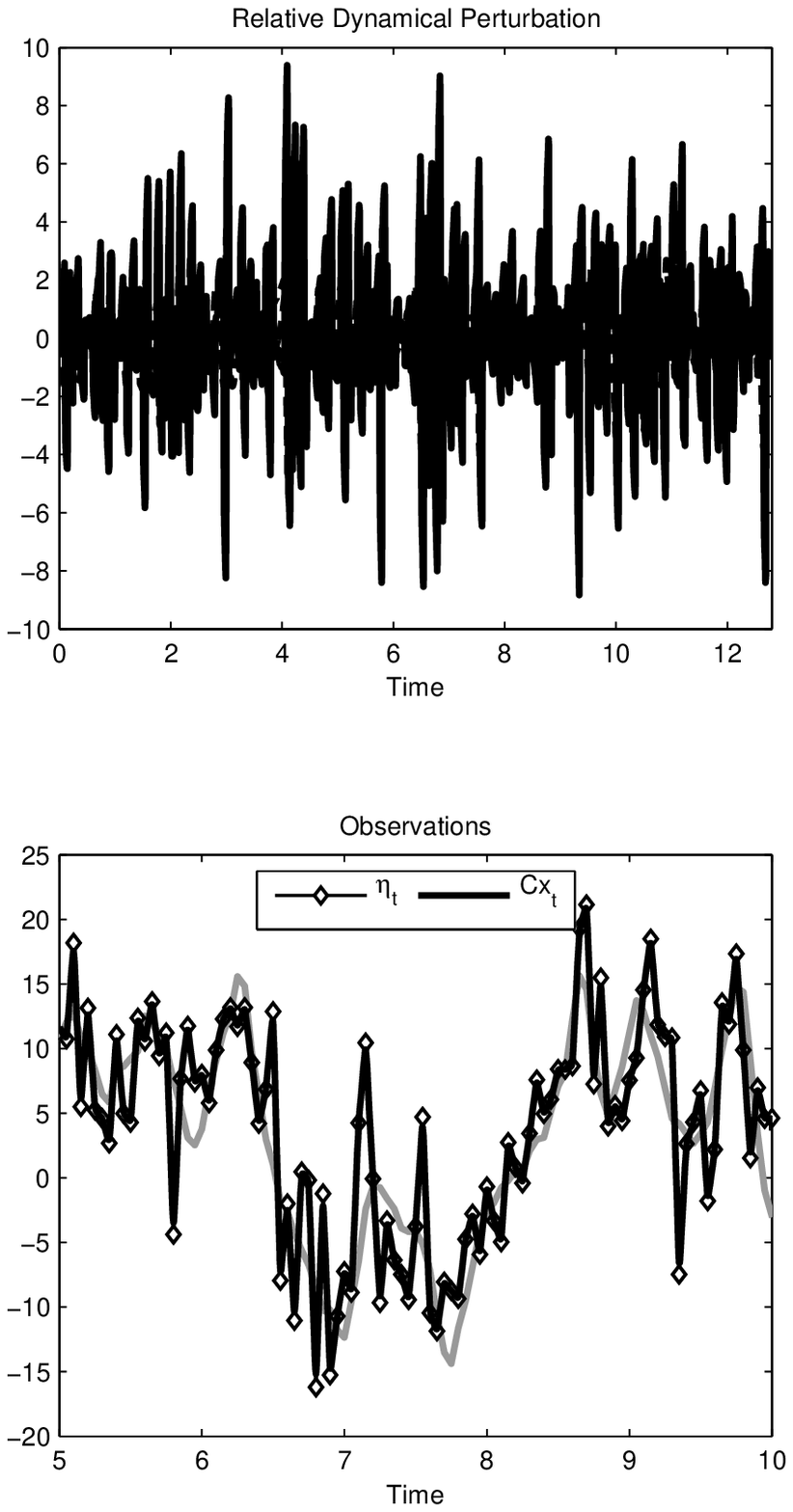, width = .49\textwidth}
\end{center}
\caption{\label{fig:am8-traj}%
Trajectories and dynamical perturbations for Lorenz'63 experiment with $\SNR_D = 5\dB, \SNR_O = 5\dB$, and $\alpha = 3.35 \cdot 10^{-4}$.
Panels on the left hand side contain the three components of the original trajectory (grey) and the assimilated trajectory (black).
Right hand side, upper panel, contains the dynamical perturbations over time. 
Right hand side, lower panel, contains observations (black narrow line with diamonds), output $y_t = Cx_t$ (black wide line), and noise--free observations $C X_t$ (grey wide line).
In this particular case, observations and output are almost identical.
This implies though that the observational noise is assimilated into the solution, leading to deviations between the assimilated and true trajectories
(left hand side panels).
Furthermore, the dynamical perturbations are large compare to the vector field (right hand side, upper panel).
}
\end{figure}
\begin{figure}
\begin{center}
\epsfig{file = 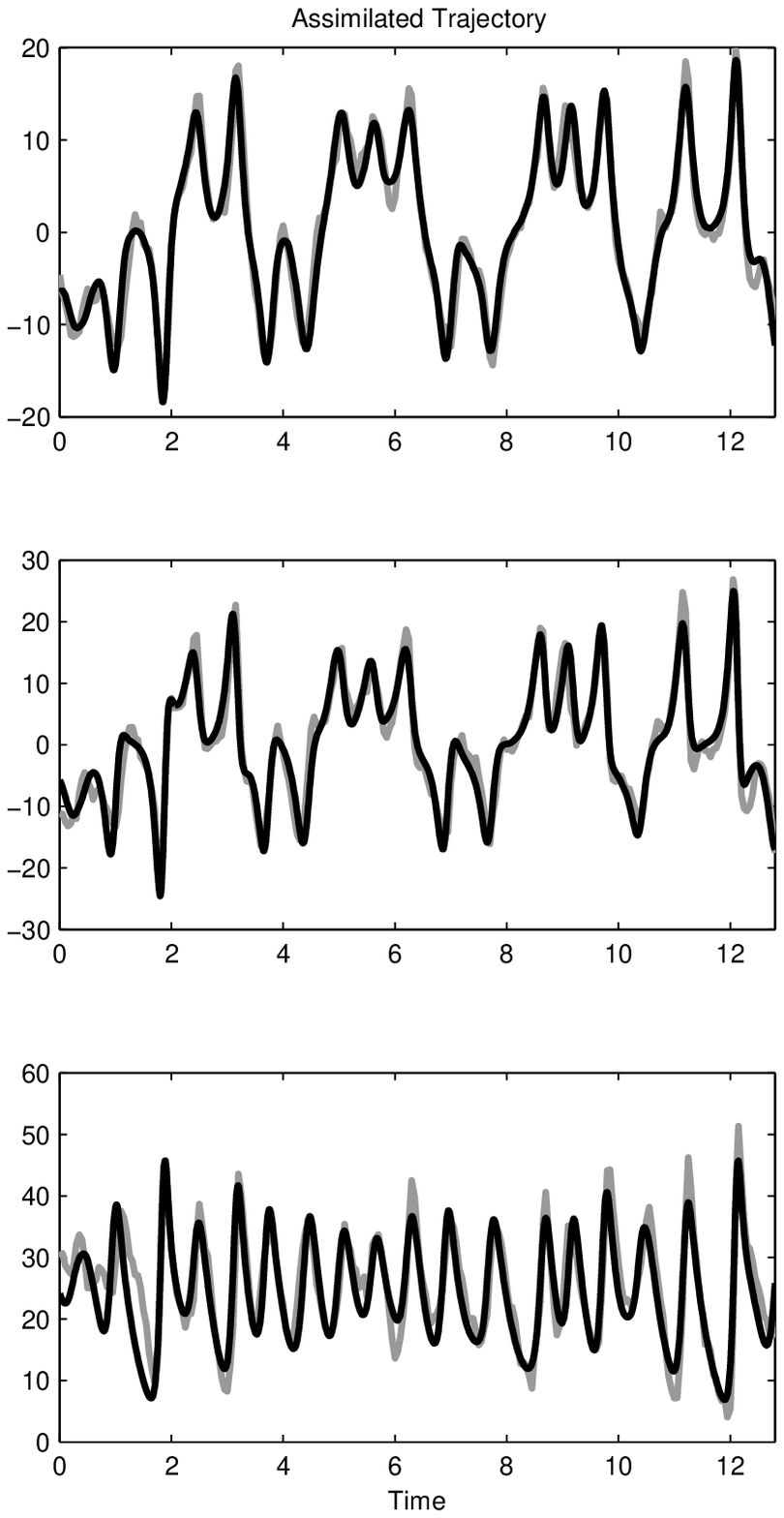, width = .49\textwidth}
\hfill
\epsfig{file = 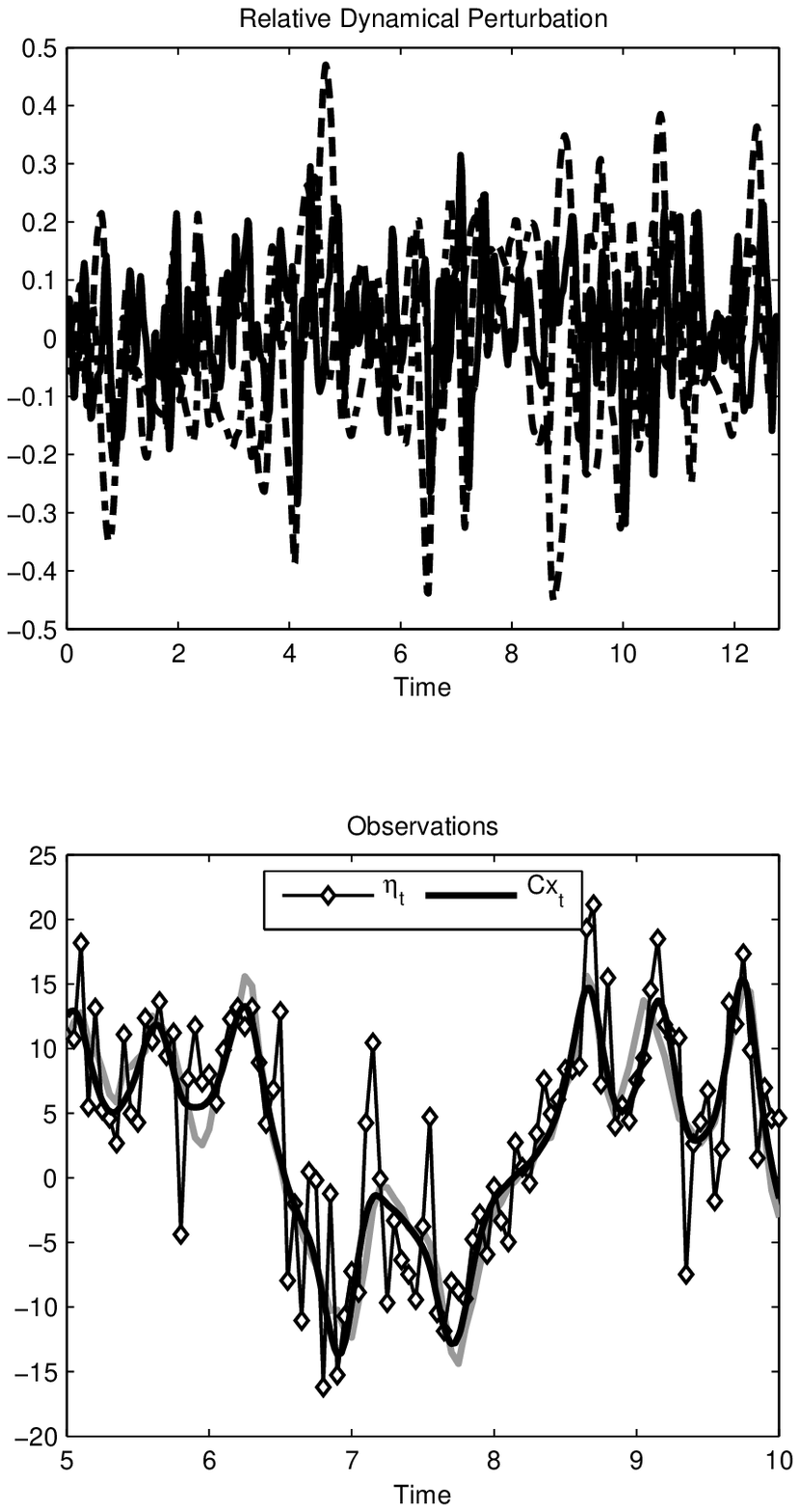, width = .49\textwidth}
\end{center}
\caption{\label{fig:a0-traj}%
Trajectories and dynamical perturbations for Lorenz'63 experiment with $\SNR_D = 5\dB, \SNR_O = 5\dB$, and $\alpha = 0.5$.
Interpretation of the plot symbols as in Figure~\ref{fig:am8-traj}.
For this value of $\alpha$, the assimilation error $A_A$ was minimum.
The output closely follows the noise--free observations, and the assimilated trajectories agree well with the true trajectories (left hand side panels).
Furthermore, the dynamical perturbations are small compared to the vector field (right hand side, upper panel).
}
\end{figure}
\begin{figure}
\begin{center}
\epsfig{file = 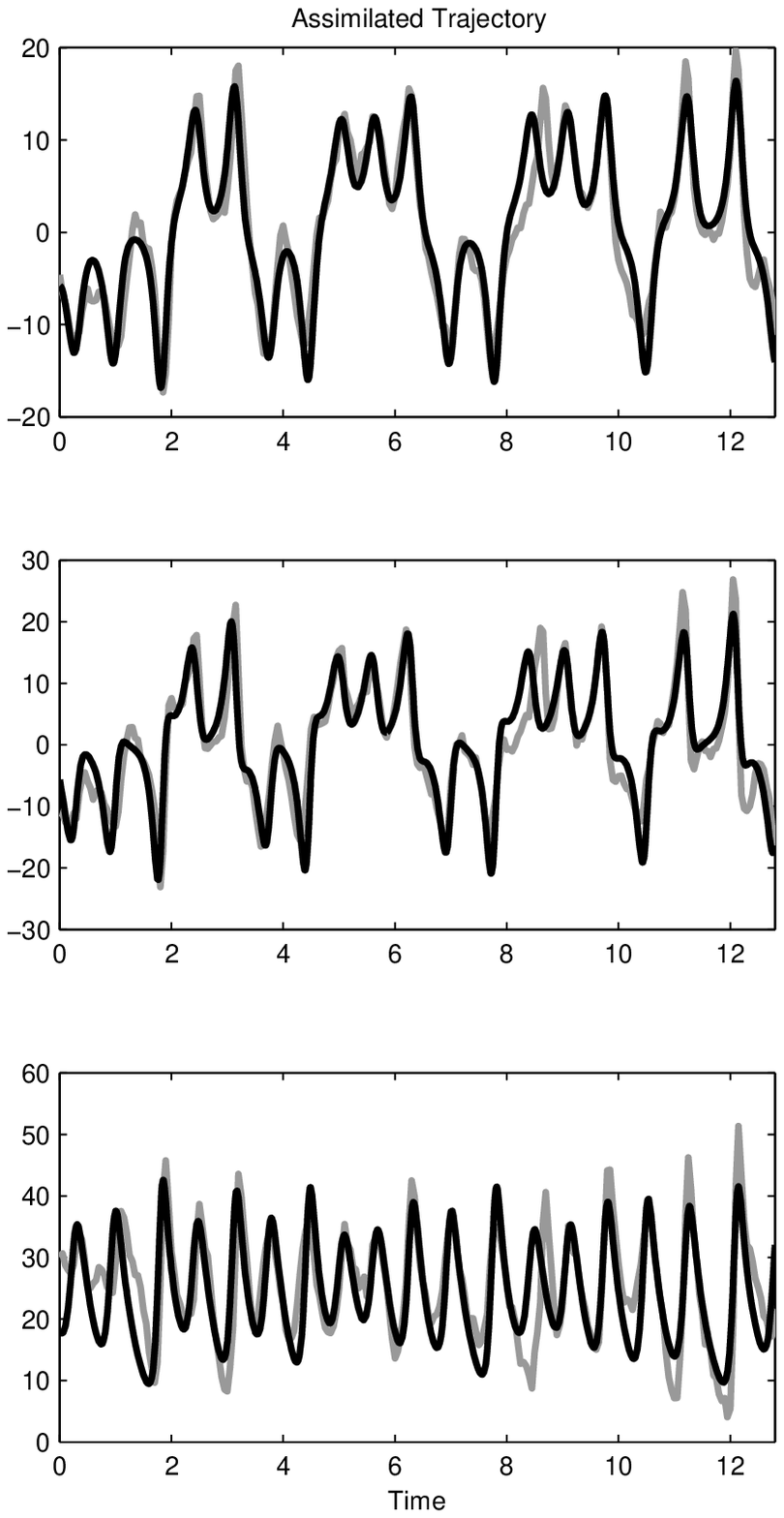, width = .49\textwidth}
\hfill
\epsfig{file = 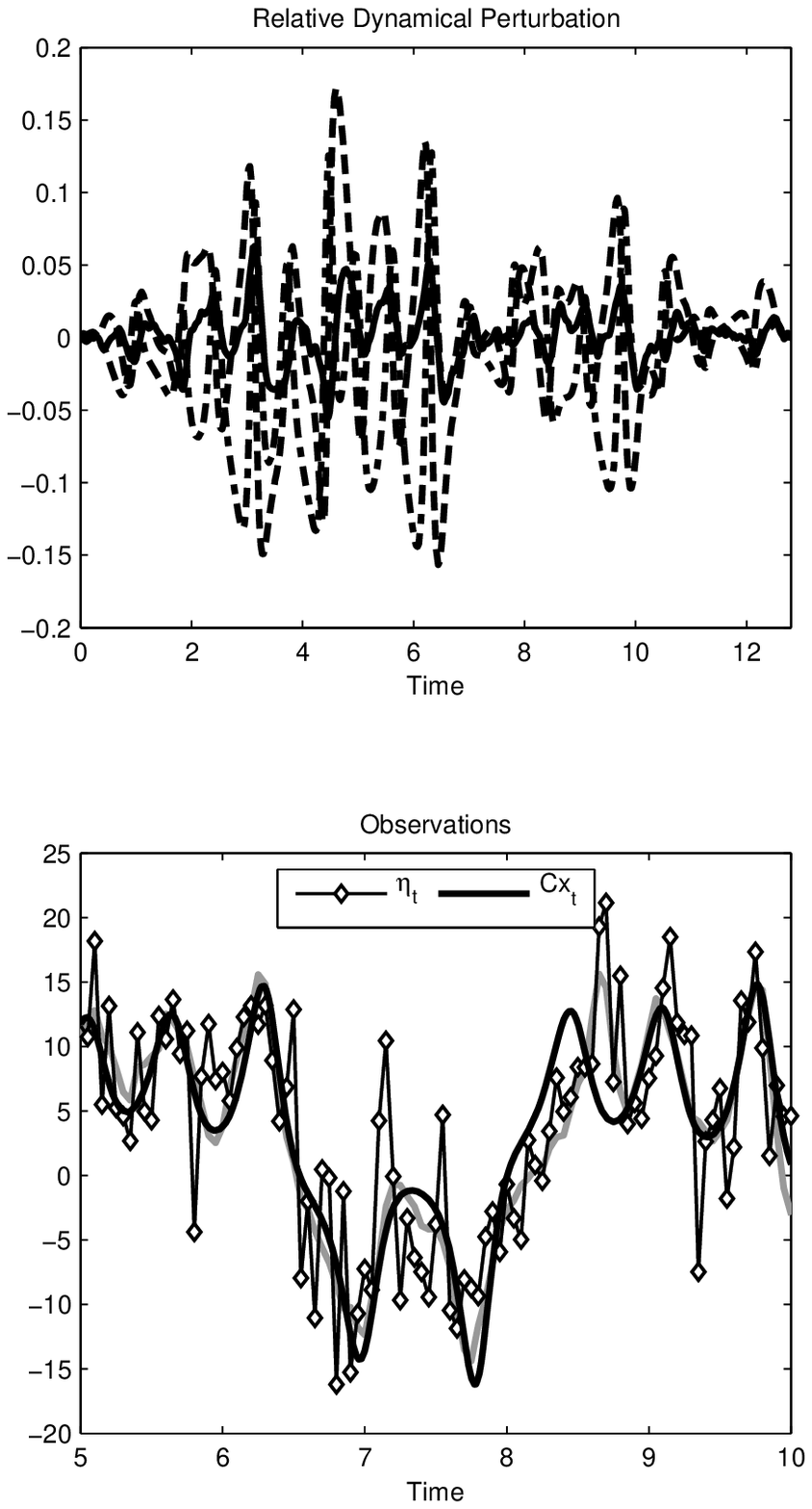, width = .49\textwidth}
\end{center}
\caption{\label{fig:a3p5-traj}%
Trajectories and dynamical perturbations for Lorenz'63 experiment with $\SNR_D = 5\dB, \SNR_O = 5\dB$, and $\alpha = 0.97$.
Interpretation of the plot symbols as in Figure~\ref{fig:am8-traj}.
The output still follows the noise--free observations, but some error in the form of phase slip is visible at around $t = 8.5$.
The assimilated trajectories still agree to some extent with the true trajectories (left hand side panels), but the phase slip at around $t = 8.5$ is visible here, too.
The dynamical perturbations are very small compared to the vector field (right hand side, upper panel), as expected.
}
\end{figure}
\begin{figure}
\begin{center}
\epsfig{file = 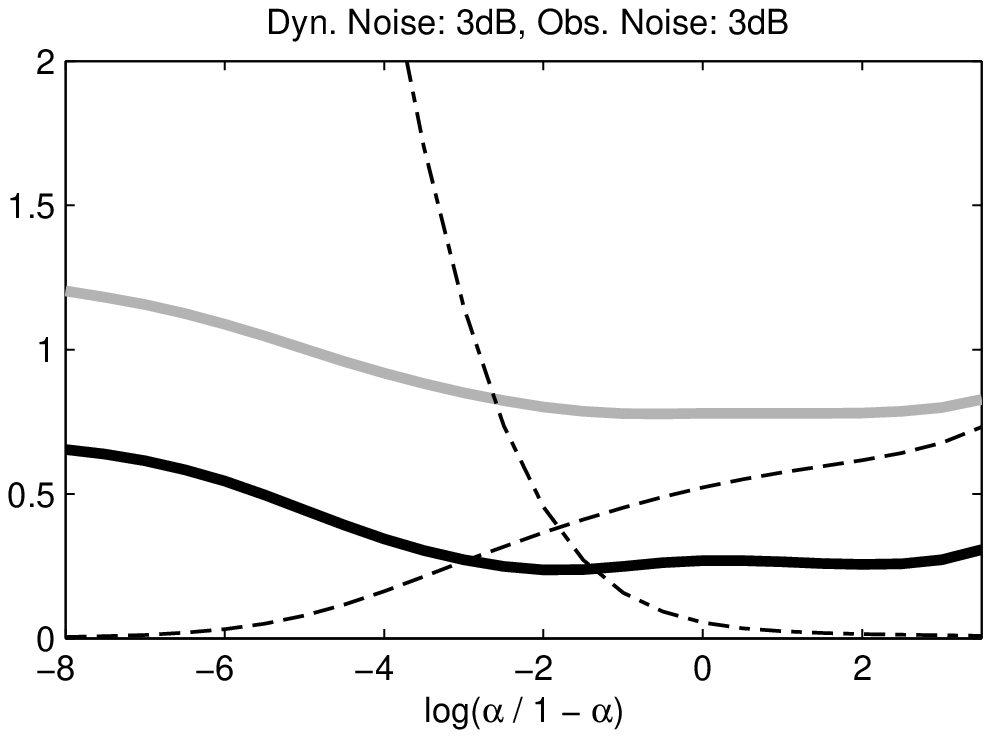}
\end{center}
\caption{\label{fig:A-dd3-do3}%
Tracking error $A_T$ (dashed line), modeling error $A_M$ (dash--dotted line), assimilation error $A_A$ (solid black line), and the assimilation error of the observed part of the state $A_1$ as a function of $\alpha$.
Results for $\SNR_D = 3\dB, \SNR_O = 3\dB$ are shown.
}
\end{figure}
\begin{figure}
\begin{center}
\epsfig{file = 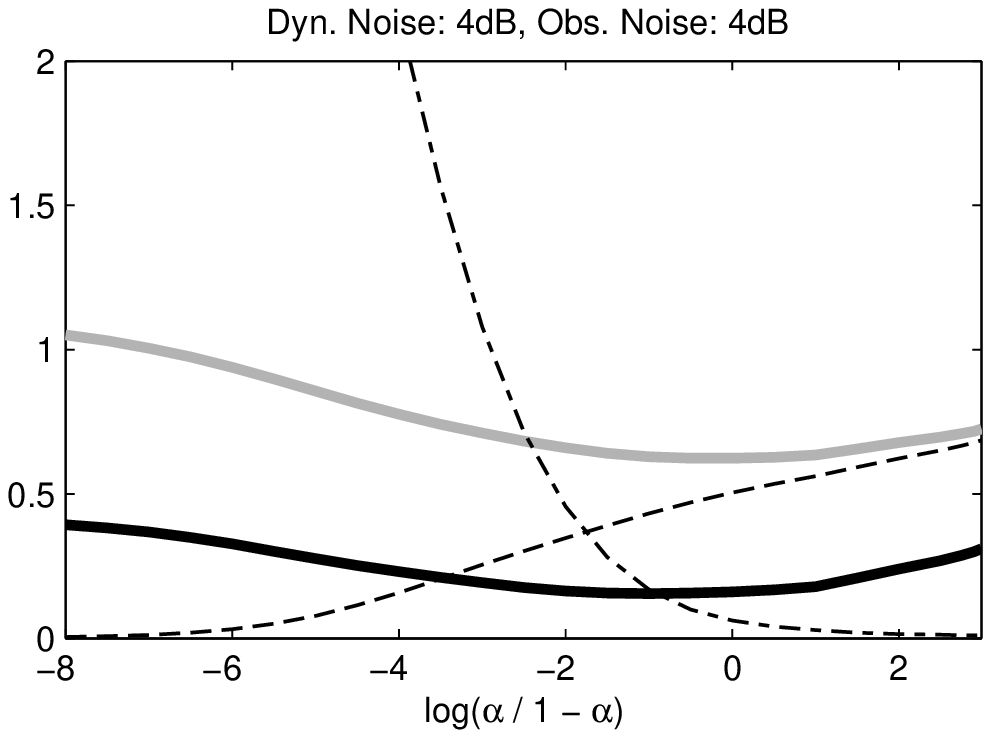}
\end{center}
\caption{\label{fig:A-dd4-do4}%
Tracking error $A_T$ (dashed line), modeling error $A_M$ (dash--dotted line), assimilation error $A_A$ (solid black line), and the assimilation error of the observed part of the state $A_1$ as a function of $\alpha$.
Results for $\SNR_D = 4\dB, \SNR_O = 4\dB$ are shown.
}
\end{figure}
\begin{figure}
\begin{center}
\epsfig{file = 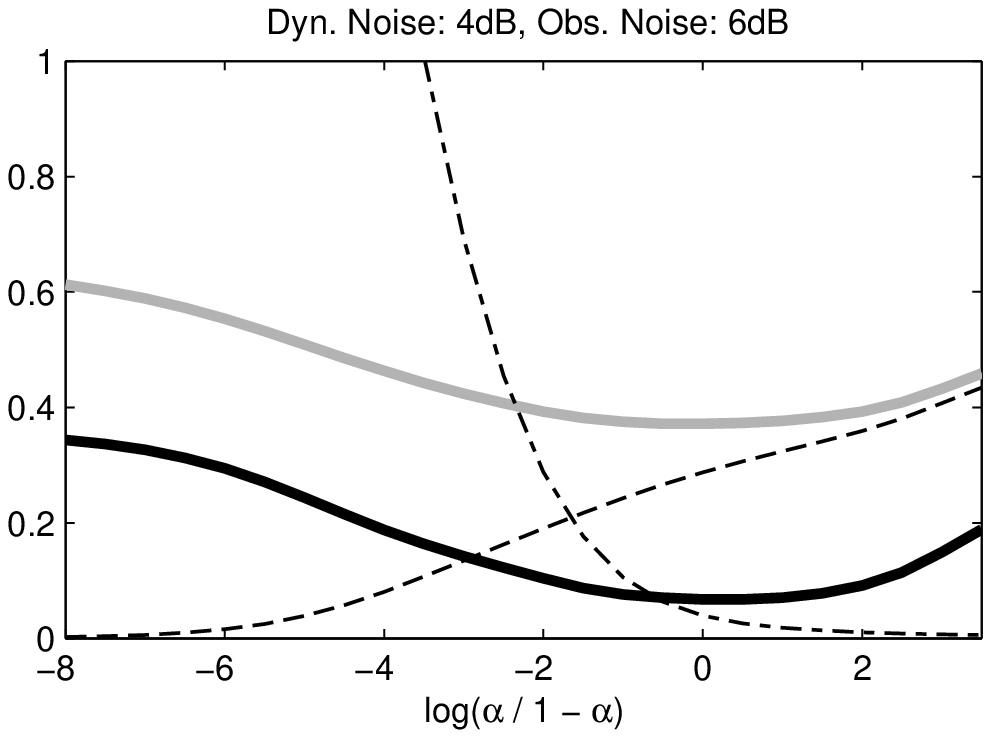}
\end{center}
\caption{\label{fig:A-dd4-do6}%
Tracking error $A_T$ (dashed line), modeling error $A_M$ (dash--dotted line), assimilation error $A_A$ (solid black line), and the assimilation error of the observed part of the state $A_1$ as a function of $\alpha$.
Results for $\SNR_D = 4\dB, \SNR_O = 6\dB$ are shown.
}
\end{figure}
\begin{figure}
\begin{center}
\epsfig{file = 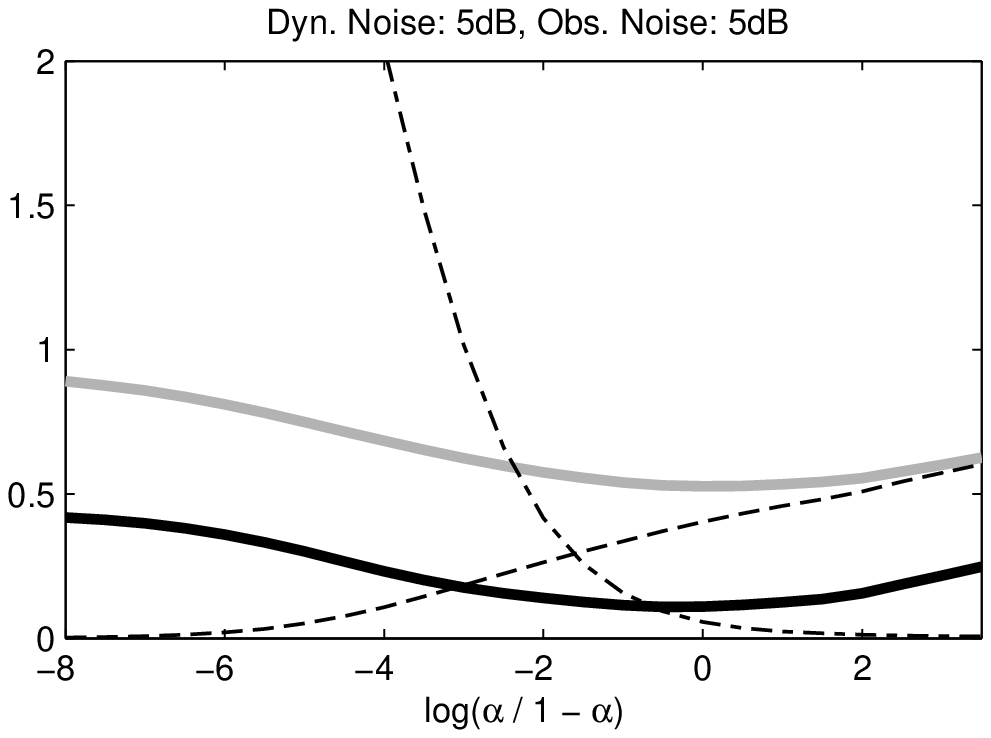}
\end{center}
\caption{\label{fig:A-dd5-do5}%
Tracking error $A_T$ (dashed line), modeling error $A_M$ (dash--dotted line), assimilation error $A_A$ (solid black line), and the assimilation error of the observed part of the state $A_1$ as a function of $\alpha$.
Results for $\SNR_D = 5\dB, \SNR_O = 5\dB$ are shown.
}
\end{figure}
\begin{figure}
\begin{center}
\epsfig{file = 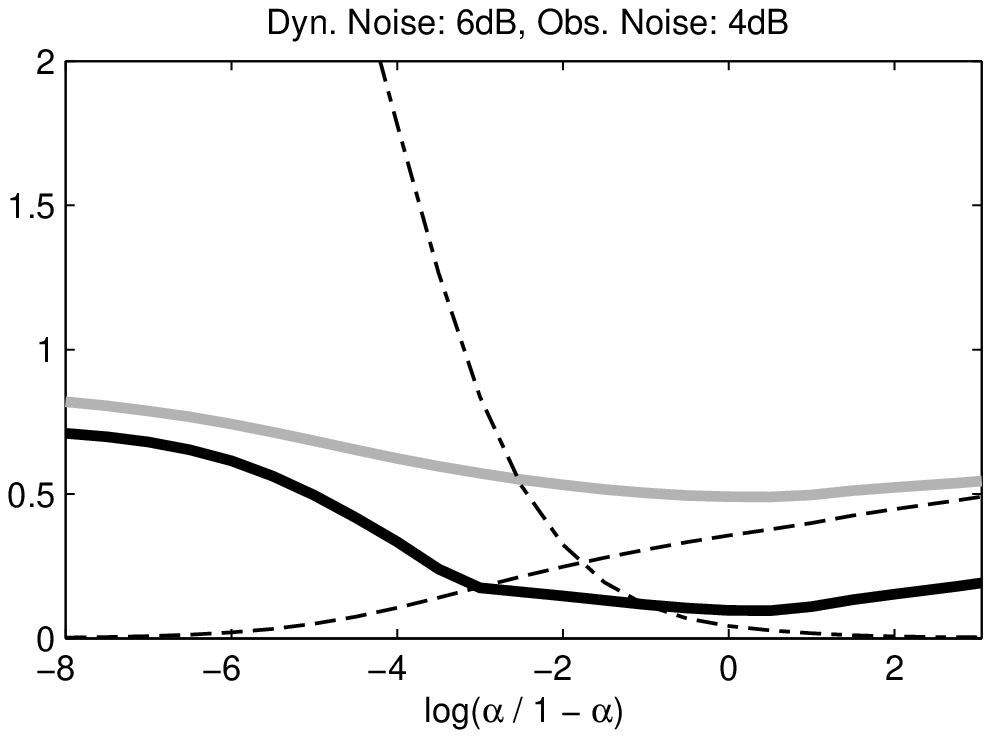}
\end{center}
\caption{\label{fig:A-dd6-do4}%
Tracking error $A_T$ (dashed line), modeling error $A_M$ (dash--dotted line), assimilation error $A_A$ (solid black line), and the assimilation error of the observed part of the state $A_1$ as a function of $\alpha$.
Results for $\SNR_D = 6\dB, \SNR_O = 4\dB$ are shown.
}
\end{figure}
\begin{figure}
\begin{center}
\epsfig{file = 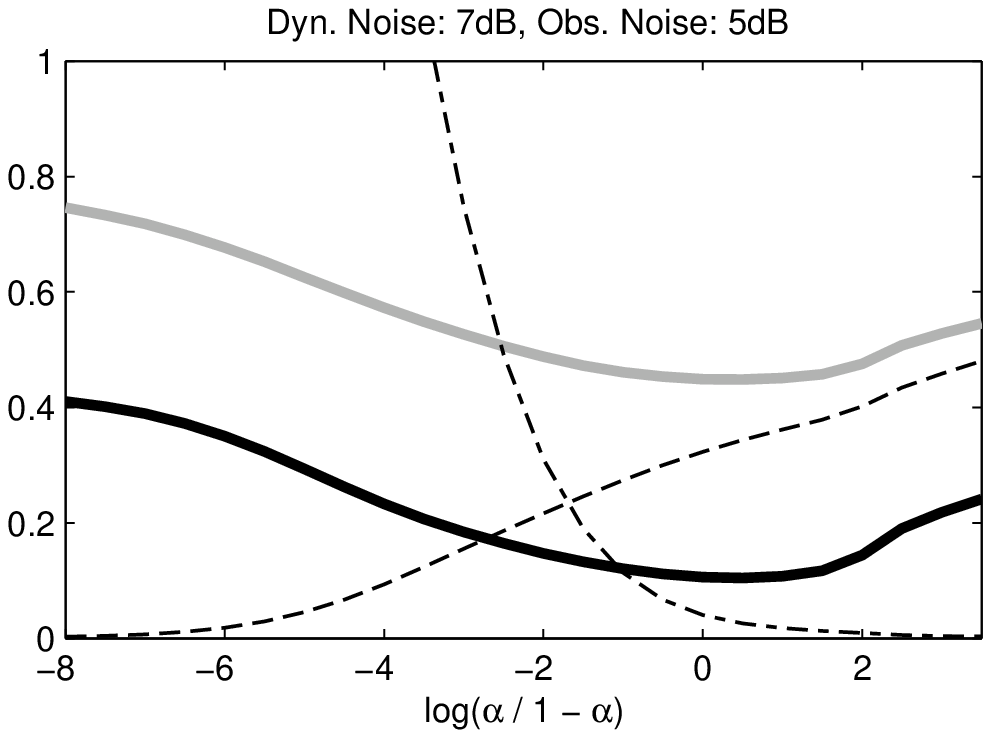}
\end{center}
\caption{\label{fig:A-dd7-do5}%
Tracking error $A_T$ (dashed line), modeling error $A_M$ (dash--dotted line), assimilation error $A_A$ (solid black line), and the assimilation error of the observed part of the state $A_1$ as a function of $\alpha$.
Results for $\SNR_D = 7\dB, \SNR_O = 5\dB$ are shown.
}
\end{figure}
\begin{figure}
\begin{center}
\epsfig{file = 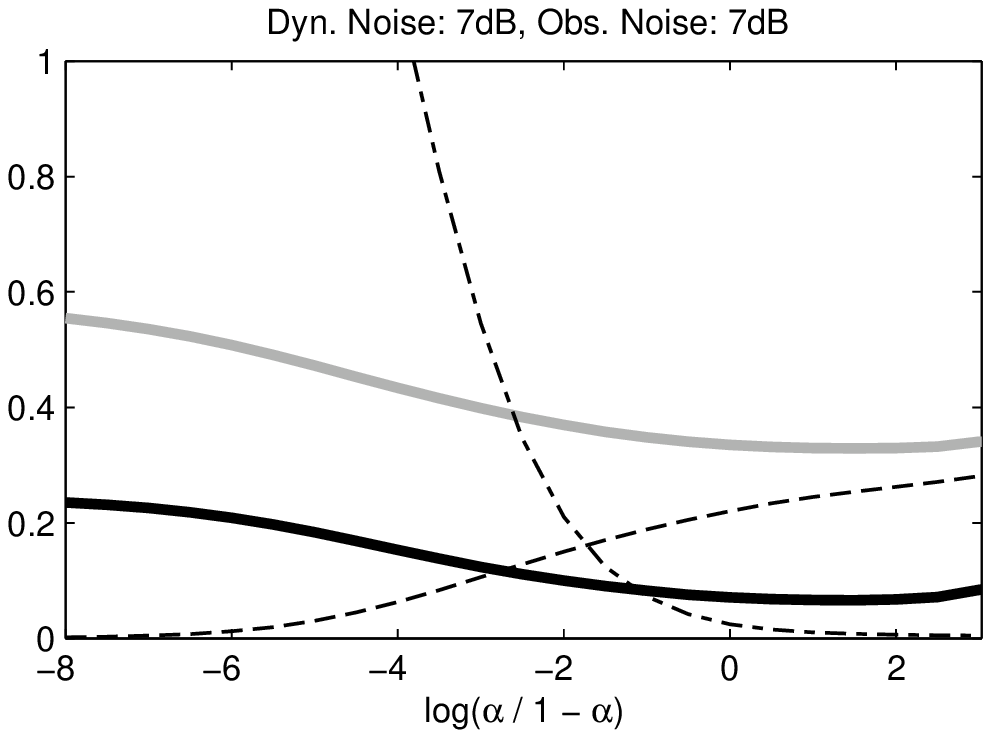}
\end{center}
\caption{\label{fig:A-dd7-do7}%
Tracking error $A_T$ (dashed line), modeling error $A_M$ (dash--dotted line), assimilation error $A_A$ (solid black line), and the assimilation error of the observed part of the state $A_1$ as a function of $\alpha$.
Results for $\SNR_D = 7\dB, \SNR_O = 7\dB$ are shown.
}
\end{figure}
\begin{figure}
\begin{center}
\epsfig{file = 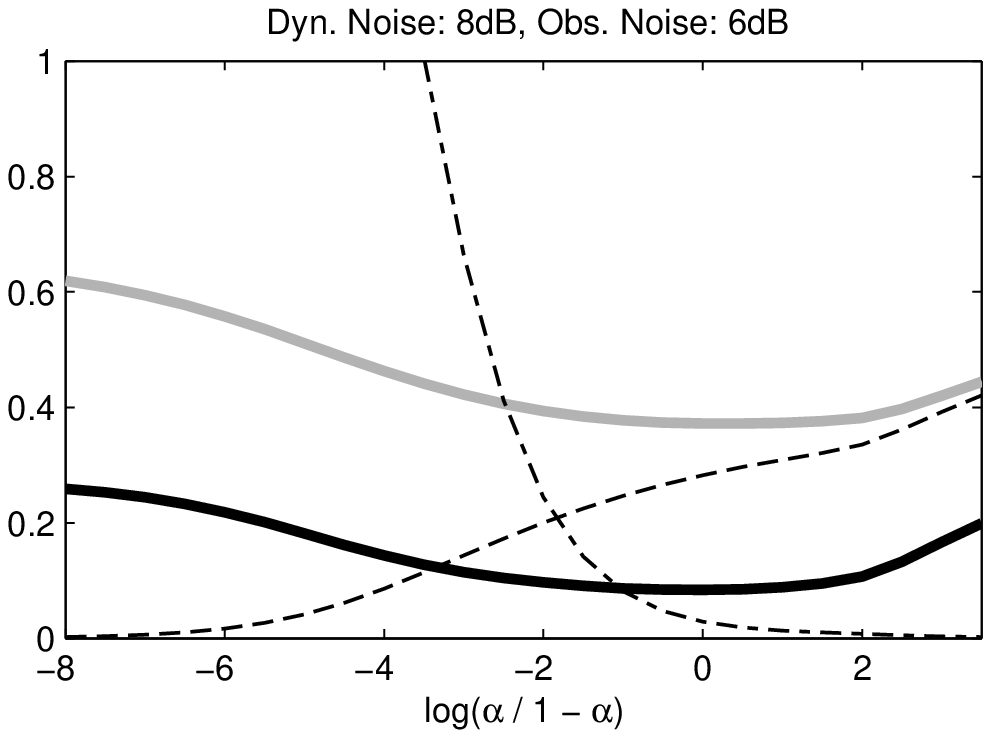}
\end{center}
\caption{\label{fig:A-dd8-do6}%
Tracking error $A_T$ (dashed line), modeling error $A_M$ (dash--dotted line), assimilation error $A_A$ (solid black line), and the assimilation error of the observed part of the state $A_1$ as a function of $\alpha$.
Results for $\SNR_D = 8\dB, \SNR_O = 6\dB$ are shown.
}
\end{figure}
\begin{figure}
\begin{center}
\epsfig{file = 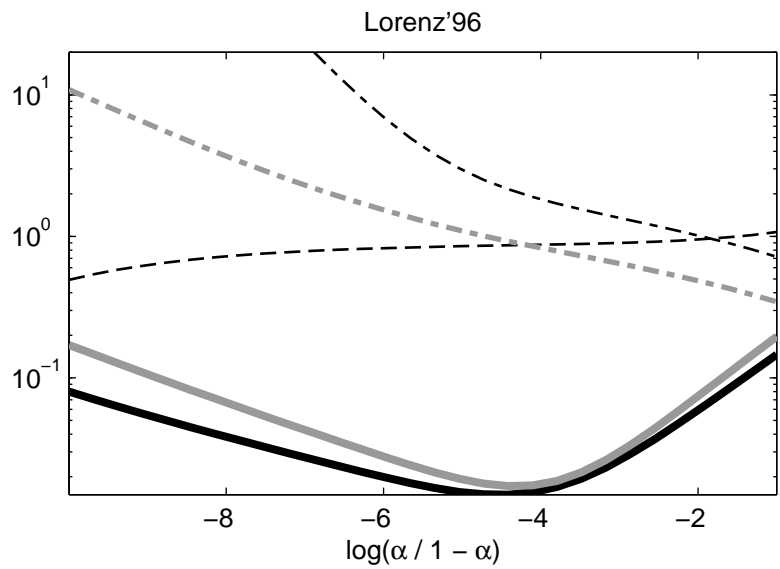}
\end{center}
\caption{\label{fig:lor96}%
This figure is to be interpreted as Figures~\ref{fig:A-dd3-do3}--\ref{fig:A-dd8-do6}  for the Lorenz'63 case.
Tracking error $A_T$ (dashed line), modeling error $A_M$ (dash--dotted line), assimilation error $A_A$ (solid black line), and the assimilation error of the observed part of the state $A_1$ as a function of $\alpha$.
Additionally, the modeling error in the hidden degrees of freedom was calculated, and is represented by the grey dash--dotted line.
Due to the wider variations of the quantities in the present case, the $y$--axis was scaled logarithmically.
}
\end{figure}